\documentclass[12pt]{reportj}
\usepackage{deluxetablej}
\usepackage{hyperref} 
\usepackage{amsmath} 
\usepackage[]{natbib}
\usepackage{amssymb}
\makeatletter
\def\@to{to}
\makeatother
\usepackage{times}
\usepackage{graphicx}
\usepackage{xcolor}
\usepackage{tocloft}
\usepackage{fancyheadings}
\emergencystretch=\maxdimen
\usepackage{datetime}
\newdateformat{ddmonthyyyy}{\THEDAY\ \monthname[\THEMONTH]\ \THEYEAR}

\renewcommand\thesection{\arabic{section}}
\renewcommand\thesubsection{\thesection.\arabic{subsection}}

\def\ssection#1{\setcounter{subsection}{0} \refstepcounter{section} \section*{\hbox to \hsize{\large\bf \arabic{section}. #1\hfill }}\label{sec} \addcontentsline{toc}{section}{\arabic{section}. #1}}
\def\ssubsection#1{\setcounter{subsubsection}{0} \refstepcounter{subsection}\subsection*{\hbox to \hsize{\normalsize\bfseries\itshape \arabic{section}.\arabic{subsection} #1\hfill}}\label{subsec} \addcontentsline{toc}{subsection}{\arabic{section}.\arabic{subsection} #1}}
\def\ssubsubsection#1{\refstepcounter{subsubsection}\subsection*{\hbox to \hsize{\normalsize\it \arabic{section}.\arabic{subsection}.\arabic{subsubsection} #1\hfill}}\label{subsubsec} \addcontentsline{toc}{subsubsection}{\arabic{section}.\arabic{subsection}.\arabic{subsubsection} #1}}

\def\ssectionstar#1{\section*{\hbox to \hsize{\large\bf #1\hfill}} \addcontentsline{toc}{section}{#1}}
\def\ssubsectionstar#1{\subsection*{\hbox to \hsize{\normalsize\bfseries\itshape #1\hfill}} \addcontentsline{toc}{subsection}{#1}}
\def\ssubsubsectionstar#1{\subsection*{\hbox to \hsize{\normalsize\it  #1\hfill}} \addcontentsline{toc}{subsection}{#1}}

\renewcommand{\cftaftertoctitle}{%
\mbox{}\hfill{\normalfont Page}}
\def\appsection#1{%
  \setcounter{subsection}{0}%
  \refstepcounter{section}%
  \section*{\hbox to \hsize{\large\bf Appendix \Alph{section}. #1\hfill}}%
  \addcontentsline{toc}{section}{Appendix \Alph{section}. #1}}

\defcitealias{dressel2007}{STIS ISR 2007-03}
\defcitealias{Ward-Duong2022}{STIS ISR 2022-01}
\defcitealias{friedman2005}{STIS ISR 2005-03}
\defcitealias{welty2018}{STIS ISR 2018-04}
\defcitealias{welty2025}{STIS ISR 2025-01}
\defcitealias{mingozzi2026}{STIS ISR 2026-03}
\defcitealias{Pascucci2011}{STIS ISR 2011-01}
\defcitealias{Sonnentrucker2015}{STIS ISR 2015-02 }
\defcitealias{hulbert1996}{STIS ISR 2026-03}
\defcitealias{hodge1998}{STIS ISR 98-12}
\defcitealias{Hodge1998-1}{STIS ISR 98-10}
\defcitealias{stisihb}{STIS Instrument Handbook}
\defcitealias{stisdhb}{STIS Data Handbook}

\begin{document}

~\\

\vspace{-2.4cm}
\noindent\includegraphics*[width=0.295\linewidth]{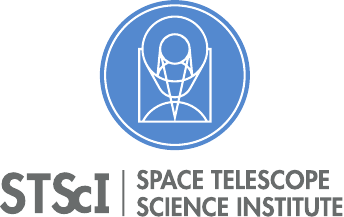}

\vspace{-0.4cm}

\begin{flushright}
    {\bf Instrument Science Report STIS 2026-04}
    
    \vspace{1.1cm}
    
    {\bf\Huge Wavelength Calibration Accuracy Across the STIS CCD: Pipeline Update and User Guidance}
    
    \rule{0.25\linewidth}{0.5pt}
    
    \vspace{0.5cm}
    
    Matilde Mingozzi$^1$, Robert Jedrzejewski$^2$, Matt Siebert$^2$, Dan Welty$^2$, \\ Sean Lockwood$^2$, Joleen Carlberg$^2$ \\
 \vspace{0.2cm}
    \footnotesize{$^1$AURA for ESA, Space Telescope Science Institute, 3700 San Martin Drive, Baltimore, MD 21218, USA \looseness=-2 \\}
    \footnotesize{$^2$Space Telescope Science Institute, 3700 San Martin Drive, Baltimore, MD 21218, USA\\}
    
    \vspace{0.5cm}

     \ddmonthyyyy\today 
\end{flushright}

\vspace{0.1cm}

\noindent\rule{\linewidth}{1.0pt}
\noindent{\bf A{\footnotesize BSTRACT}}

{\it \noindent 
We investigate the wavelength calibration accuracy of STIS CCD spectra as a function of detector position, focusing on the E1/E2 pseudo-apertures. Comparison between measured lamp line centroids and laboratory wavelengths shows that, while the standard \texttt{calstis} solution is stable at the detector center, significant offsets are present toward the CCD edges and increase with time. 
We used a test version of \texttt{calstis} in which the wavelength calibration step employs a row-selected cross-correlation, leading to improved wavelength calibration accuracy for edge extractions. 
Motivated by these results, we implemented an update to \texttt{calstis4} that applies a row-selected wavecal procedure to observations taken with the E1/E2 pseudo-apertures, while preserving the standard procedure for nominal extractions.
The updated pipeline was then used to reprocess the entire STIS archive in MAST.
Validation tests confirm improved agreement between nominal and E1/E2 spectra without significant side effects for more than 97\% of the datasets in the MAST archive. The main exceptions are the majority of G230MB and a few G230LB and G430M datasets, where the lower signal-to-noise at the CCD edge makes the lamp lines harder to detect and cosmic-ray residuals can dominate the cross-correlation, leading to incorrect wavelength shifts.
A Jupyter Notebook is also provided to enable similar corrections in case the cross-correlation fails, for extractions at positions other than E1/E2, or for multiple extractions along the slit. Finally, we discuss the potential impact of the wavelength calibration accuracy on science results.}

\vspace{-0.1cm}
\noindent\rule{\linewidth}{1.0pt}

\renewcommand{\cftaftertoctitle}{\thispagestyle{fancy}}
\tableofcontents



\vspace{-0.3cm}
\section{Introduction}\label{sec:Introduction} 

The STIS CCD detector shows a gradual rotation over time, reflected in measurable changes in spectral trace angles \citepalias[$\sim 0.0031-0.0041$~deg/yr; ][]{dressel2007}, in the long-term evolution of the CCD flat fields \citepalias[$\sim 0.0031$~deg/yr; ][]{Ward-Duong2022}, and in independent astrometric analyses of archival imaging tied to Gaia DR2 \citep[$\sim 0.0038$~deg/yr; ][]{nguyen2021}.
In particular, \citetalias{dressel2007} quantified the rotation angle of the spectral traces for a subset of CCD settings (i.e., G230LB/2375, G430L/4300, G750L/7751, and G750M/6768,6581,8561), and stored it in the DEGPERYR column of the one-dimensional Spectrum Trace Table (SPTRCTAB; \texttt{qa31608go\_1dt.fits}) introduced in 2006. This information is used by the \texttt{calstis} pipeline to correct the cross-dispersion shift of the trace position. However, this remains limited to a subset of CCD settings. Furthermore, no corresponding correction has been established for the dispersion shift up to now.

Analyses over the years on CCD dispersion-monitoring programs have shown that the absolute scale (or zero-point) of the CCD dispersion solutions at the nominal position (center of the CCD, row~$\sim $~512) is accurate to within $\sim0.2$~pixels, and unchanged from before the Hubble Servicing Mission 4 (SM4; May 2009), when STIS was successfully repaired and resumed operations \citepalias{Pascucci2011, Sonnentrucker2015, welty2018}. 
However, zero point offsets are known to be larger at the edges of the CCD \citepalias{friedman2005,welty2018,mingozzi2026} and to degrade significantly over time \citepalias{mingozzi2026}.
\citetalias{friedman2005} examined the accuracy achieved at pseudo-apertures E1 and E2, which place the target spectrum closer to the CCD output amplifier (row $\sim$~900) to mitigate CCD charge transfer inefficiency, from Cycle 11 and Cycle 12 (pre-SM4) dispersion monitor spectra. In particular, they identified systematic wavelength offsets at row $\sim$~900 that were generally within the 0.2~pixel accuracy requirement, except for G750L/7751, for which the mean wavelength calibration error was estimated to be $\approx 0.37$~pixel.
More recently, \citetalias{mingozzi2026} quantified dispersion shifts at different CCD positions over two decades (Cycles 11–33) for 14 CCD grating settings by cross-correlating dispersion monitor spectra obtained at the same detector location with the corresponding Cycle 11 spectra used as the reference (see their Figures~4–11). 
They found comparable temporal trends across all settings, with dispersion-axis shifts increasing progressively over time from the detector center toward both CCD edges and reaching $\gtrsim0.5$--$1.5$~pixels by Cycle~33.
Also, they derived rotation rates for each grating and central-wavelength configuration (consistent with previous measurements when available; see their Tables~1–3), concluding that the observed wavelength shifts are driven by CCD rotation. \looseness=-1

The dependence of the CCD rotation’s impact on wavelength calibration accuracy with detector position arises from the way the calibration procedure has historically been implemented in the \texttt{calstis} pipeline \citepalias{hulbert1996,hodge1998}.
In particular, the wavelength calibration \texttt{calstis} step measures spectral (SHIFTA1) and spatial (SHIFTA2) shifts on the detector - the zero-point offsets - caused by the Mode Select Mechanism (MSM) positioning uncertainties and thermal drifts within STIS. 
This is achieved by cross-correlating one or more Pt/Cr-Ne line lamp exposures (``wavecal" exposures) - acquired with the MSM in the same configuration as the science data (see \citetalias{welty2018} Table~1) - with a reference lamp spectrum obtained pre-launch. 
By default, this cross-correlation is optimized for the nominal position, even for observations using E1/E2 pseudo-apertures. 
Given the widespread use of the E1 pseudo-aperture (accounting for $\sim26$\% of STIS CCD long-slit exposure time between 2000 and 2026; see Appendix~\ref{app:e1-usage}), it is important to improve the wavelength calibration accuracy at E1, bringing it to a level comparable to that achieved at the nominal position.

In this ISR, we show how the wavelength calibration accuracy varies across the CCD detector (Section~\ref{sec:comp-friedman}) and describe the steps leading to the most recent update to the \texttt{calstis} pipeline (\texttt{version 3.5.0}), implemented to correct for the CCD rotation effect at the E1/E2 position (Sections~\ref{sec:cal-acc} and \ref{sec:new-pipeline}).
In Section~\ref{sec:new-pipeline}, we also present validation tests of the updated pipeline using CCD dispersion-solution monitoring data obtained over Cycles 11–33, along with other science programs' datasets, and give guidance to users.
Finally, in Appendix~\ref{app:wavecal-for-users} we describe a new Jupyter Notebook that corrects for wavelength calibration accuracy offsets when the source spectrum is located neither at the nominal nor at the E1 position, or in the case of extended sources, where spectra extracted at multiple CCD positions must be corrected consistently.



\lhead{}
\rhead{}
\cfoot{\rm {\hspace{-1.9cm} Instrument Science Report STIS 2026-04 Page \thepage}}


\section{The Degrading of Wavelength Calibration Accuracy at the Edges of the CCD}\label{sec:comp-friedman}
To test the wavelength calibration accuracy at different positions on the CCD (including nominal and E1 apertures) and as a function of time (to account for CCD rotation), we used data from the CCD dispersion solution monitor programs\footnote{This link shows all STIS calibration programs per cycle: \url{https://www.stsci.edu/hst/instrumentation/stis/calibration}} 
from Cycle 33 to Cycle 7, which have been monitoring a large set of CCD gratings and central wavelengths (i.e., G230LB/2375, G430L/4300, G750L/7751, G230MB/2416,2697,3315, G430M/3165,3680,4961,5471, G750M/6768,6581,8561). 
In particular, prior to Cycle 11 the LINE lamp was used, whereas from Cycle 11 onward the HITM1 lamp was consistently adopted, enabling high signal-to-noise spectra even at the edges of the CCD.
In this section, we focus on the G750L/7751 and G430L/4300 settings, which have exhibited larger wavelength inaccuracies than the other configurations since \citetalias{friedman2005}. 

\begin{figure}[!h]
    \centering
    \includegraphics[width=.49\linewidth]{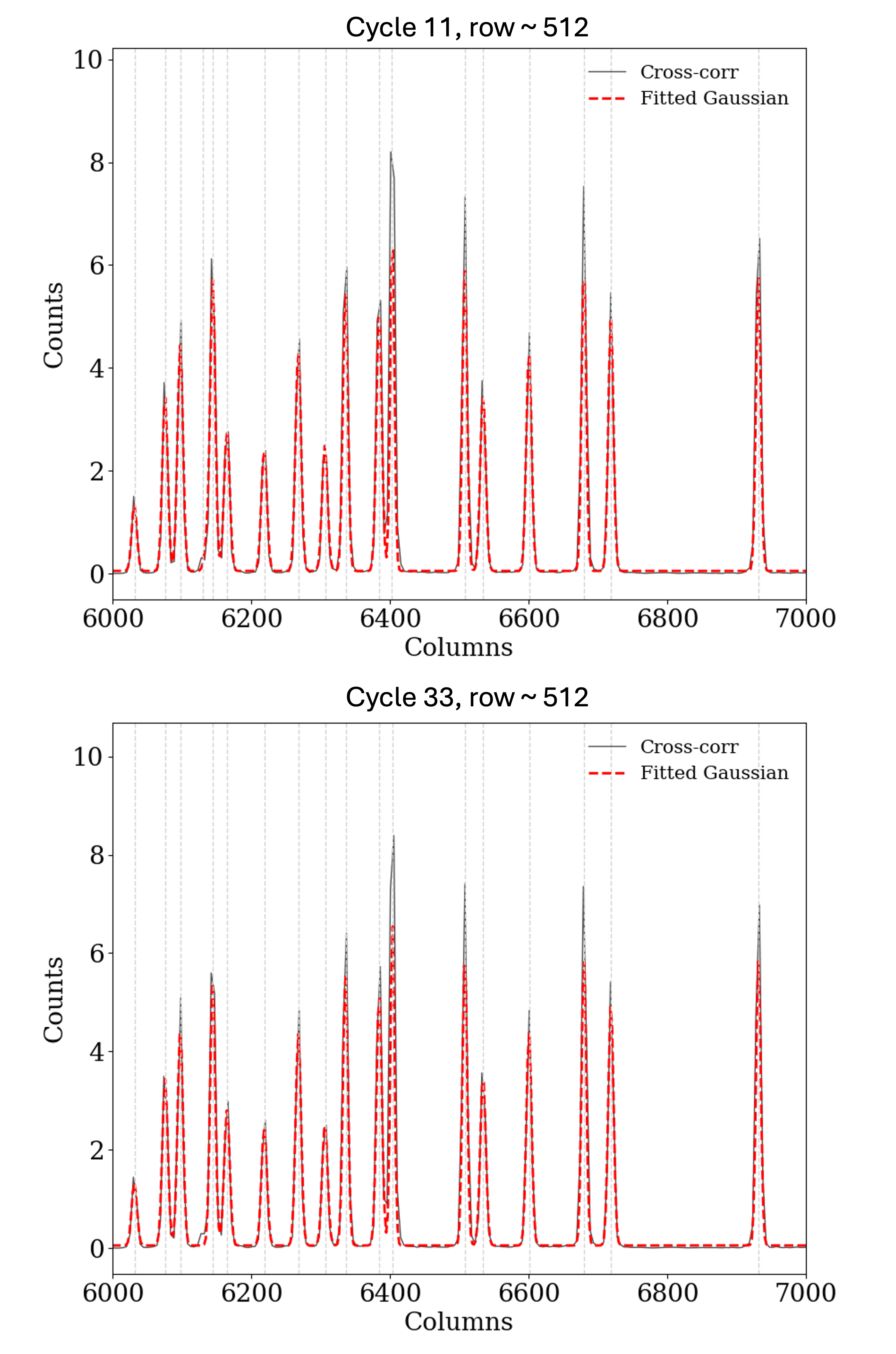}
        \includegraphics[width=.49\linewidth]{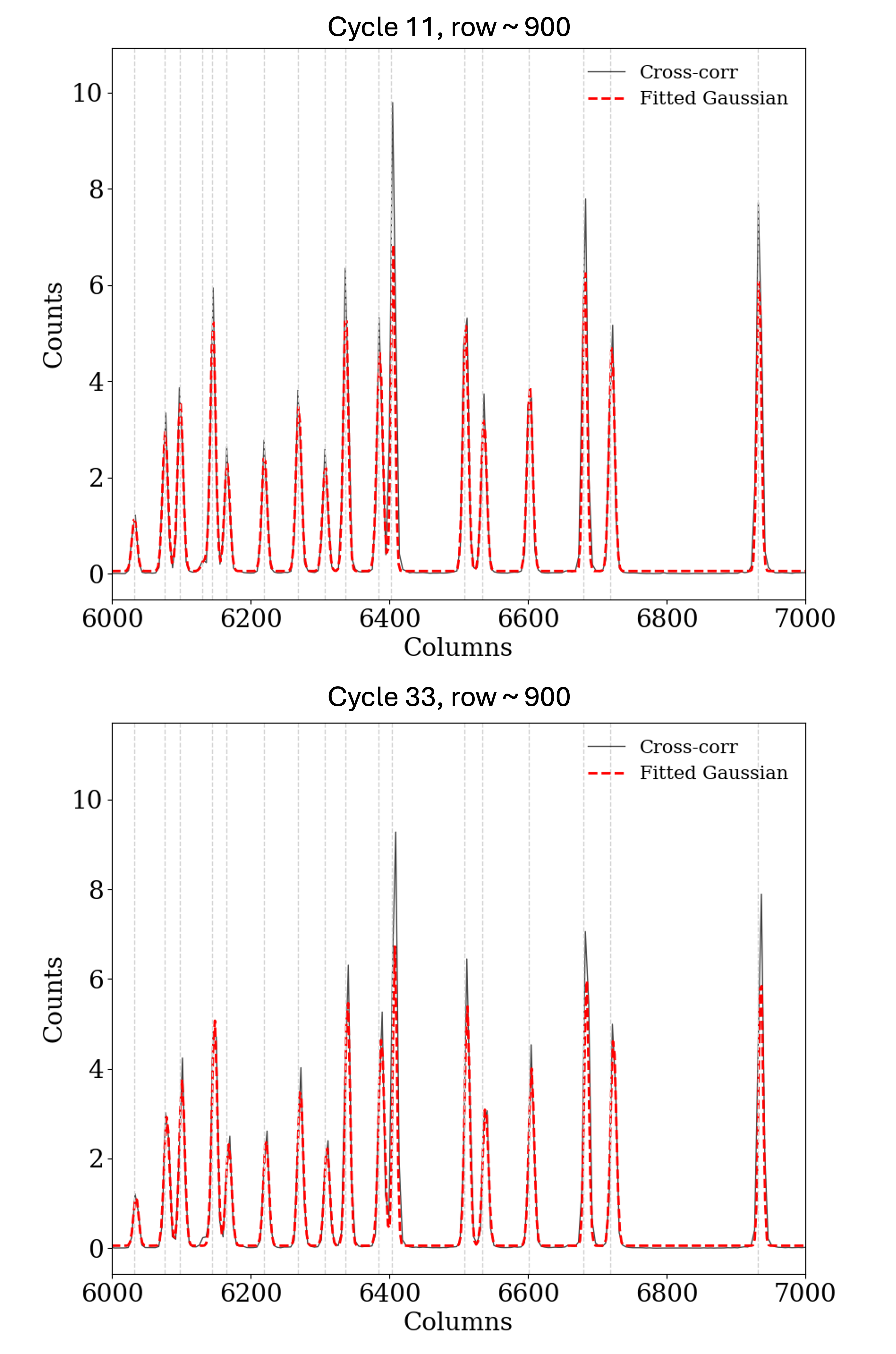}
      \caption{Example of the fitting procedure for G750L/7751 spectra extracted at the detector center in Cycle 11 (top left) and Cycle 33 (bottom left), and at the E1 position in Cycle 11 (top right) and Cycle 33 (bottom right). Vertical lines mark the reference laboratory wavelengths. The fitted Gaussian centroids agree well with the reference line positions at the detector center in both cycles, while a clear wavelength shift is visible at E1 in Cycle 33, as shown by the offset between the fitted centroids and the reference lines.}
         \label{fig:gaussianfitting-example}
   \end{figure}

To make the best comparison as possible with \citetalias{friedman2005}'s results, we reproduced the same analysis step by step (see their Section Data Analysis), first calibrating the raw lamp files using \texttt{calstis}, and then extracting from the calibrated flat-fielded output images (\texttt{\_flt.fits}) lamp files 32 spectra, centred in different columns, from 16 to 1008, using the \texttt{stistools} task \texttt{x1d} (see \citetalias{mingozzi2026} Figure~1). 
We defined a list of reference laboratory lines by identifying the main peaks (above a chosen threshold) in the central spectrum for each Cycle and matching them to NIST\footnote{\url{https://physics.nist.gov/PhysRefData/ASD/lines_form.html}} emission lines to derive the corresponding wavelengths. We then selected specific subranges to include only bright, unblended lines, excluding weaker or blended features that would otherwise affect the results.
Finally, we fitted the identified emission lines in all 32 extracted spectra for each Cycle with single Gaussian profiles, as shown in Figure~\ref{fig:gaussianfitting-example} for G750L/7751 (Cycle 11 vs.\ Cycle 33 and center vs.\ E1), initializing the centroids using the reference laboratory line list. 
Inspection of Figure~\ref{fig:gaussianfitting-example} shows consistency at the center between Cycle 11 and 33, and a clear wavelength shift at E1, as indicated by the offsets between the fitted Gaussian centroids and the reference line positions.

Comparing the fitted line centroids with the corresponding laboratory wavelengths as a function of detector position and cycle allows us to quantify the wavelength calibration accuracy across the CCD and over time.
\begin{figure*}[!h]
\centering
  \includegraphics[width=1.\linewidth]{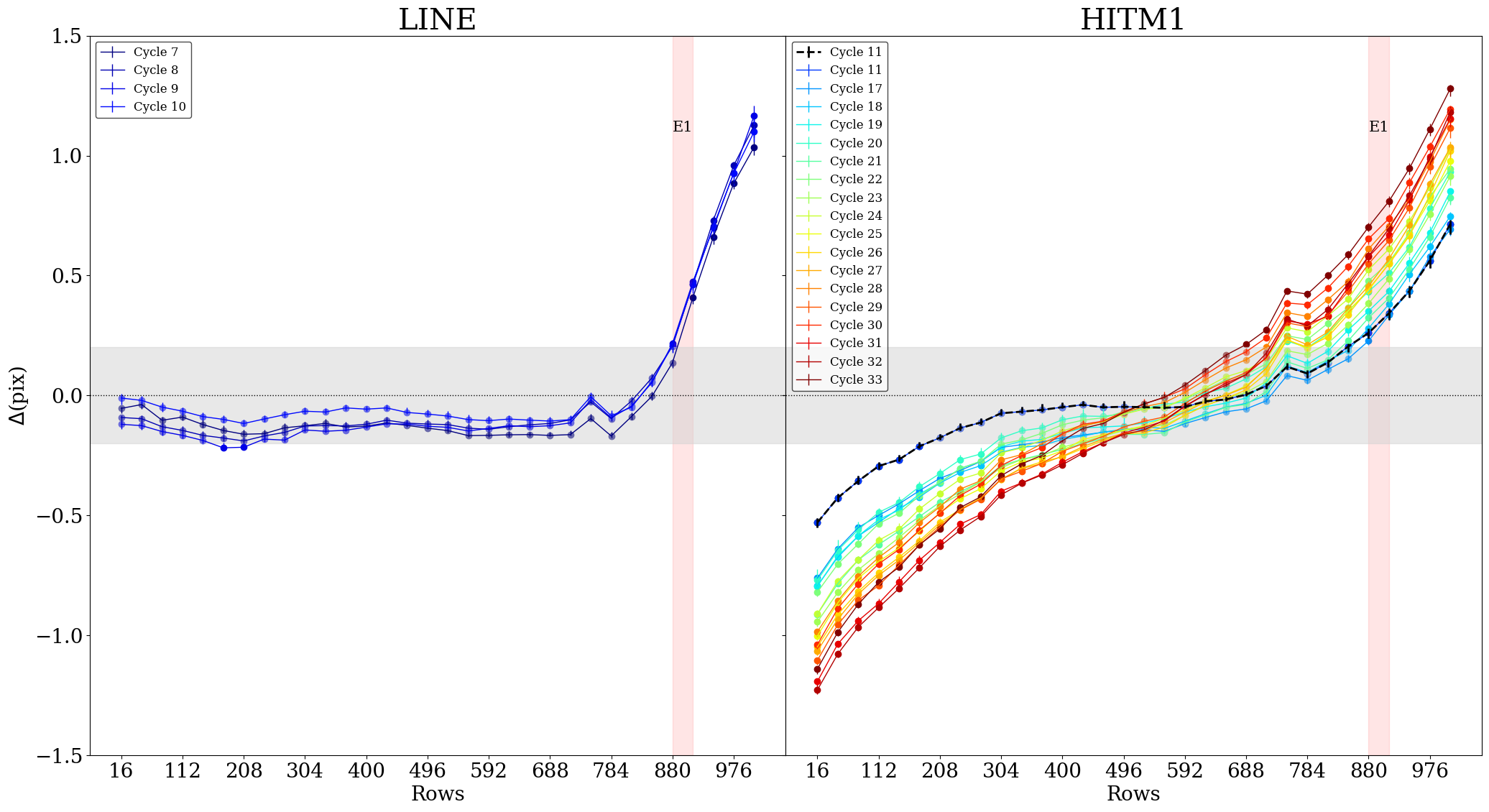}
  \includegraphics[width=1.\linewidth]{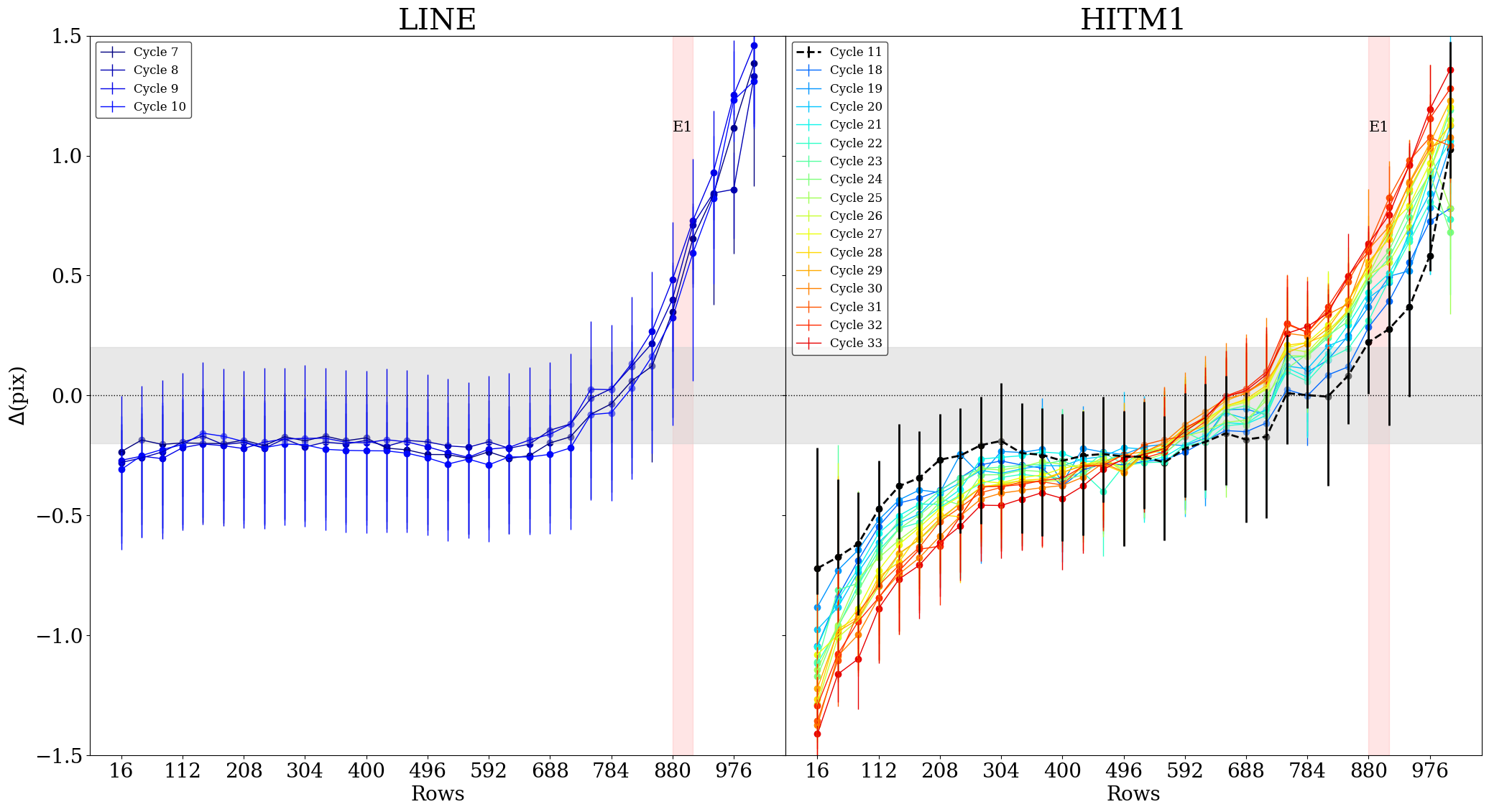}
  \caption{Upper panels: Median shift (in pixels) of LINE (left) and HITM1 (right) lamp-line centroids relative to laboratory wavelengths at different CCD positions for the G750L/7751 setting. The E1 position (row $\sim$900) is marked in red, and the horizontal shaded region indicates the $\pm$0.2 pixel accuracy requirement. Cycle 11 results are shown in black and are consistent with \citetalias{friedman2005}. Error bars represent the 16th and 84th percentiles divided by the square root of the number of lines used at each position. The shift increases with time, as indicated by the color-coding by Cycle, and is attributed to CCD rotation (see also \citetalias{mingozzi2026}). Lower panels: Same for G430L/4300.} 
\label{fig:friedman-comparison}
\end{figure*}
Figure~\ref{fig:friedman-comparison} shows that the median difference between the fitted line centroids and the laboratory wavelengths remains broadly consistent across cycles at the center of the CCD, in agreement with previous studies \citepalias{Pascucci2011, Sonnentrucker2015, welty2018}. 
In contrast, an offset that increases with time appears toward the edges of the CCD, as clearly shown by the HITM1 lamp results (Figure~\ref{fig:friedman-comparison}, right panels). 
The similar temporal trend observed for both G750L and G430L\footnote{Consistently with \citetalias{friedman2005}, the difference between the measured line centroids and the laboratory wavelengths at the CCD center for G430L is not centered at zero, but shows a small negative offset.} (as well as the consistent behavior found for all the monitored CCD settings in \citetalias{mingozzi2026}) reinforces that this effect is mainly driven by the CCD rotation discussed in Section~\ref{sec:Introduction} \citepalias{dressel2007,Ward-Duong2022}. 

We note that the offset is mostly visible on one side of the CCD for the LINE lamp results (Figure~\ref{fig:friedman-comparison}, left panels), possibly due to the low signal-to-noise at the CCD top edge. However, it is not possible to assess the temporal behavior of this effect, as LINE lamp observations were not obtained in recent cycles of dispersion monitor programs. 
Overall, the presence of a systematic shift at the detector edge even in cycles prior to Cycle~11 indicates that additional effects beyond CCD rotation may be contributing to the observed offsets.

In Cycle 33, the shift between nominal and E1 emission line centroids shown in Figure~\ref{fig:friedman-comparison} is approximately 0.9 pixels and 1 pixel for G750L and G430L, respectively, corresponding to about 4.4~\AA\ and 2.7~\AA, assuming dispersions of $\sim$4.9~\AA\,pixel$^{-1}$ for G750L and $\sim$2.7~\AA\,pixel$^{-1}$ for G430L. 
However, when considering the full CCD extent, the shifts can reach up to approximately 5.4~\AA\ for G750L and 3.8~\AA\ for G430L. 
These inaccuracies can have a significant impact on the interpretation of science observations (see Section~\ref{sec:impact-on-science}). In Section~\ref{sec:cal-acc}, we therefore describe the approach adopted to address this issue and develop a solution.

\subsection{Impact on Science Results}\label{sec:impact-on-science}
To estimate the potential scientific impact of the wavelength offsets discussed above, we convert the measured shifts into velocity units using
\[
\Delta v = c\,\Delta\lambda/\lambda_{\rm obs}.
\]
The aim is to illustrate the systematic effect that could arise if wavelength shifts in science spectra (e.g., for extended sources observed at multiple CCD positions or for point sources placed at E1) were misinterpreted as real velocity offsets rather than calibration effects. 

In particular, the maximum shifts observed at E1 for the L gratings (up to $\sim 1$~pixel from the detector center toward each edge) correspond to spurious velocity offsets up to $\approx 200$~km~s$^{-1}$. 
The effect of CCD rotation is present at a similar level (up to $\sim 0.5-1$~pixel) in pixel units for the M gratings (see \citetalias{mingozzi2026}), corresponding to spurious velocity offsets of $\approx 20$~km~s$^{-1}$ given their finer wavelength scale per pixel.

These values are smaller than the nominal spectral resolution of both the L gratings ($R \approx 500-1000$, corresponding to FWHM~$\approx 300-600$~km~s$^{-1}$) and the M gratings ($R \approx 5000-10000$, corresponding to FWHM~$\approx 30-60$~km~s$^{-1}$). However, systematic wavelength shifts dependent on detector position at the level discussed here can still significantly affect velocity measurements.
As a result, they can impact velocity-sensitive applications, including spatially resolved kinematic measurements when spectra are extracted at different positions along the slit in case of extended sources, as well as redshift measurements of point sources observed at E1.

\section{Wavelength Calibration in \texttt{calstis}: Existing Method, Limitations, and Workarounds}\label{sec:cal-acc} 
Updating the \texttt{calstis} pipeline is not trivial because of its existing architecture and design, which makes structural changes difficult. Nevertheless, limited and targeted modifications are possible.

The wavelength calibration in \texttt{calstis} is handled through a series of dedicated modules that process contemporaneous wavecal exposures and propagate the resulting spectral shifts to the science data. 
The calibration assumes the nominal dispersion relation derived from early (pre-launch and immediate post-launch) wavecal observations, which describes the dependence of wavelength on detector position and spectral order. 
Small offsets relative to this reference solution are then applied using the wavecal-derived zero-point corrections stored in the SHIFTA keywords.
Within this calibration scheme, the lines of constant wavelength are not exactly perpendicular to the dispersion (trace) direction, but include a small intrinsic tilt that produces a positional offset between the central location and E1 even in the absence of detector rotation; this effect is accounted for in the nominal dispersion relation. 
The corresponding pipeline structure is described in detail by \citetalias{hulbert1996,hodge1998,Hodge1998-1}.
The summary of the main steps relevant for wavelength calibration provided below applies to first-order spectra, as all STIS CCD spectroscopic modes operate in first order:
\begin{itemize}
    \item \texttt{calstis0} is the wrapper task that orchestrates the pipeline flow and calls the individual calibration modules according to the header switches. It controls the execution order and ensures that the appropriate calibration steps are applied to the input data.
    \item \texttt{calstis4} processes the 2-D rectified (rectification done by \texttt{calstis7}) wavecal exposures to determine shifts in the dispersion (SHIFTA1) and cross-dispersion (SHIFTA2) directions introduced by non-repeatability of the MSM and other thermal effects. 
    Using a spectrum summed over the full illuminated extent of the slit, the module cross-correlates the data with the reference lamp template (\texttt{l421050oo\_lmp.fits}) to determine the wavelength and spatial zero-point offsets. 
    These offsets are expressed in pixel units and written to the headers of temporary files.
    \item \texttt{calstis12} writes SHIFTA1 and SHIFTA2 into the science headers. These values are then used to calculate the wavelength array in the spectra extraction module (\texttt{calstis6}).
\end{itemize}

Since \texttt{calstis4} uses a summed 1-D spectrum, the SHIFTA1 value computed in the original pipeline is representative of the detector center (``nominal wavelength calibration"). 
The same correction is subsequently applied by \texttt{calstis6}, independently of the extraction position on the cross-dispersion axis on the detector. 
The results of \citetalias{friedman2005}, based on Cycle~10--11 datasets, suggested that the nominal wavelength calibration was generally adequate even for spectra extracted near the edges of the CCD detector (deviations within $\sim0.2$ pixels, apart from G750L; see Section~\ref{sec:Introduction}). 
However, subsequent monitoring (i.e., \citetalias{welty2018,mingozzi2026} and Figure~\ref{fig:friedman-comparison}) is showing that, due to the gradual rotation of the CCD, these offsets at the detector edges have been increasing over time. 
As a result, the nominal wavelength calibration is no longer adequate for spectra extracted near the edges of the CCD.
On the other hand, a row-selected cross-correlation is expected to provide improved results.

The work described in Section~\ref{sec:test-pipeline} 
motivated both a pipeline update that improves wavelength calibration accuracy at the E1 position (see Section~\ref{sec:new-pipeline}) and the development of a Jupyter Notebook (described in Appendix~\ref{app:wavecal-for-users}) that applies a row-selected cross-correlation to the wavecal exposures associated with the observations, enabling the derivation of row-specific offsets to be applied to the science data.

\subsection{Test Wavecal Pipeline with Row-Selected Cross-Correlation}\label{sec:test-pipeline}
To assess whether a row-by-row cross-correlation could address this issue, a test version of the pipeline was developed that saves intermediate wavecal products and recomputes wavelength shifts over selected row ranges, instead of using a summed 1-D spectrum collapsed over the full spatial direction.
In this test version (not available to users), the command 
$$ 
cs0.e \,\,\, \text{-}s \,\,\, rawfile.fits
$$
saves (i.e., -s) the intermediate geometrically corrected image (rectifiedfile\texttt{\_w2d\_tmp.fits}\footnote{Note that this temporary file is a larger image than the raw image of $\approx80$ pixels, added around to absorb shifts \citepalias{hodge1998}. This difference needs to be accounted for when defining the row range to compute the cross-correlation in this test version of the pipeline (not available to users).}), while 
$$
cs4.e \,\,\, \text{-}firstrow \,\,\, \text{-}lastrow \,\,\, rectifiedfile\_w2d\_tmp.fits
$$
recalculates SHIFTA1 and SHIFTA2 using only the specified rows on the detector to create the collapsed spectrum on which the cross-correlation is performed.
    \begin{figure}[htbp]
    \centering
    \includegraphics[width=1.\linewidth]{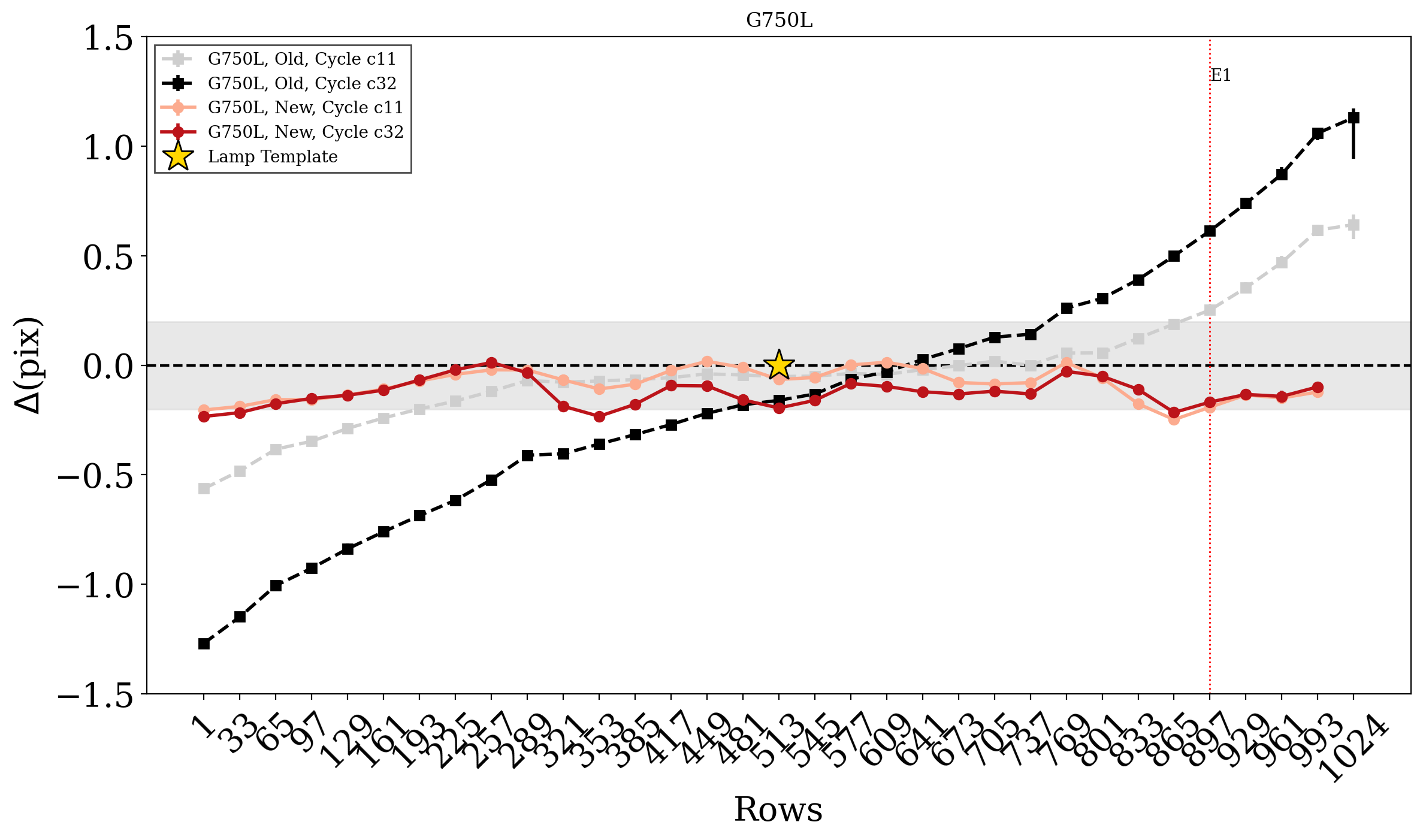}
    \includegraphics[width=1.\linewidth]{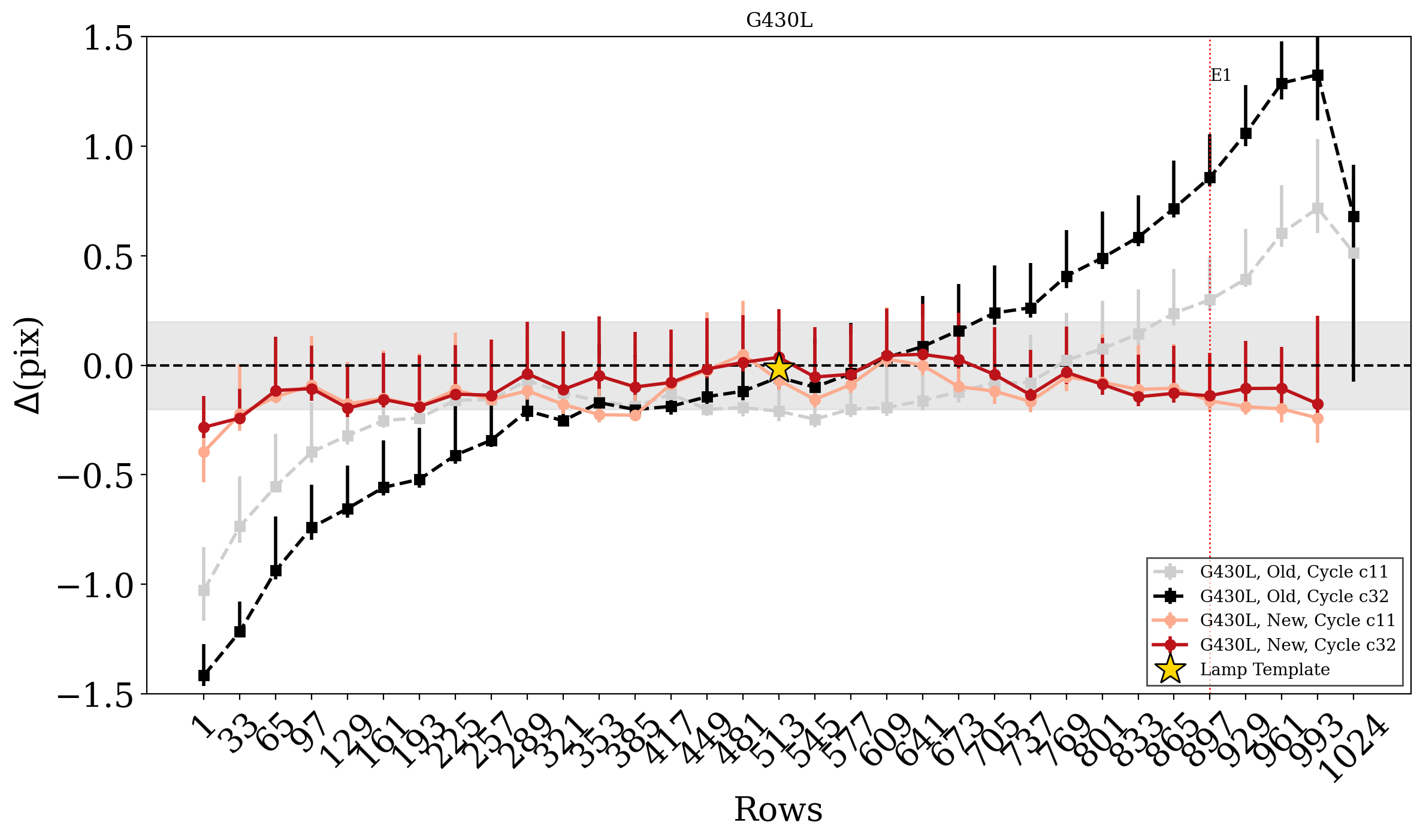}
      \caption{Upper panel: Measured median pixel shifts as a function of extraction location across the CCD using G750L/7751 wavecal data from Cycle 11 (lighter colors) and Cycle 32 (darker colors), extracted with both the current pipeline (gray/black squares) and the test pipeline (light red/red dots). The E1 position (row $\sim900$) is marked by the vertical line, while the horizontal shaded region indicates the $\pm 0.2$~pixel accuracy requirement. Bottom panel: Same for G430L/4300. This illustrates the improvement in meeting the wavelength accuracy requirement across the CCD when using a row-selected cross-correlation.}
     \label{fig:calprograms-test-pipeline}
   \end{figure}
   
We applied this test pipeline to the G750L and G430L dispersion-solution monitor data discussed in Section~\ref{sec:comp-friedman}. For each row range, we recalculated SHIFTA1 and extracted the corresponding spectrum using the \texttt{stistools.x1d} task. We then applied the Gaussian-fitting procedure described above to derive the median shifts, in pixels, between the measured line centroids and the corresponding reference wavelengths.
Figure~\ref{fig:calprograms-test-pipeline} compares the measured median pixel shifts as a function of extraction position across the CCD for G750L/7751 (upper panel) and G430L-4300 (lower panel), using the standard and test pipelines for the Cycle 11 and Cycle 32 wavelength-solution monitor data described in Section~\ref{sec:comp-friedman}. 
In particular, it shows that the wavelength calibration accuracy obtained with the test pipeline is within $\pm 0.2$~pixels across the entire CCD. The observed fluctuations and error bars mainly reflect the limitations of the Gaussian-fitting approach, which depends on the choice of emission lines and on line blending effects that vary with the grating spectral resolution.

Overall, this illustrates how row-selected cross-correlation improves agreement with the wavelength accuracy requirement across the CCD, particularly near the detector edges.
This test version of the pipeline is not available to users, as it requires manual input and does not operate within the automated pipeline framework. Questions about this approach can be directed to the \href{https://stsci.service-now.com/hst}{STIS Help Desk}.

\section{Pipeline Update for Improved Wavelength Calibration at E1}\label{sec:new-pipeline}
The updated \texttt{calstis} (\texttt{version 3.5.0}, of \texttt{hstcal version 3.2.0}) 
implements an improved wavelength calibration for CCD data obtained at the E1/E2 pseudo-aperture locations. 
In particular, when either E1 or E2 is detected from the science header (keyword {\it PROPAPER}), the corresponding detector region (approximately $40$ rows around row $\sim900$) is used to perform the cross-correlation and compute the SHIFTA1 value. 
This pipeline update affects only \texttt{calstis4}; all other modules and steps of the wavelength calibration procedure described in Section~\ref{sec:cal-acc} remain unchanged. 
If users wish to reprocess the data by running the individual modules directly instead of using \texttt{stistools}, the \texttt{-e} parameter must be specified when running \texttt{calstis4.e} to ensure that the cross-correlation is performed near the E1/E2 position (i.e., around row $\sim900$ on the detector; e.g., \texttt{calstis4.e -e rectified-file}).

The choice to implement this correction in a limited and targeted way reflects the structure of the \texttt{calstis} pipeline. In particular, the wavelength offsets are calculated before the spectrum location is determined, and therefore the pipeline does not yet have information about the exact extraction position along the slit at that stage of the processing. 
As a consequence, observations obtained with non-default target placements (e.g., POS-TARG offsets or extended sources along the slit) are not fully accounted for by the wavelength offset correction implemented in the updated pipeline. In these cases, residual wavelength shifts may partly reflect the systematic effect due to the CCD rotation. Users encountering these situations are referred to Appendix~\ref{app:wavecal-for-users} for further details when high-precision wavelength calibration is required.

\subsection{Validation Tests for the Updated Pipeline}\label{sec:validation-tests}
A first validation test was performed using targets from the SNAP program 16230 observed with G430L/4300 at both the nominal and E1 positions (9 targets in total), processing the datasets with both the original and updated versions of the pipeline starting from the raw exposures. 
For context, this program was designed to obtain NUV and blue optical spectra of $\sim200$ Magellanic Cloud OB stars in the ULLYSES sample.

\begin{figure}[!h]
    \centering
      \includegraphics[width=.9\linewidth]{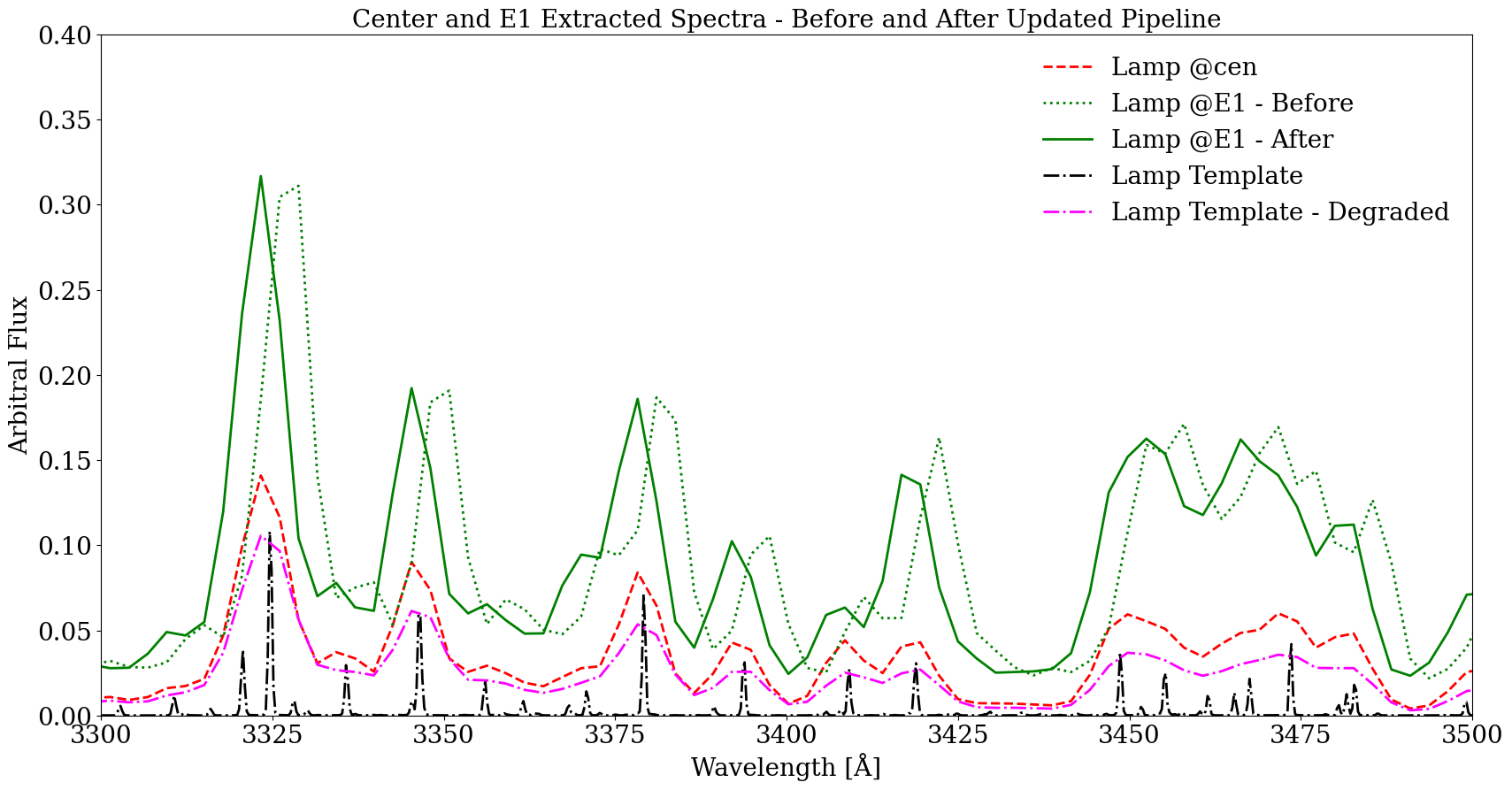}
      \caption{Comparison between the old and updated wavelength calibrations applied to G430L wavecal spectra taken for the source SK-67-207 (PID 16230).} 
         \label{fig:snaptest-wavecal}
   \end{figure}

\begin{figure}[!h]
    \centering
      \includegraphics[width=.8\linewidth]{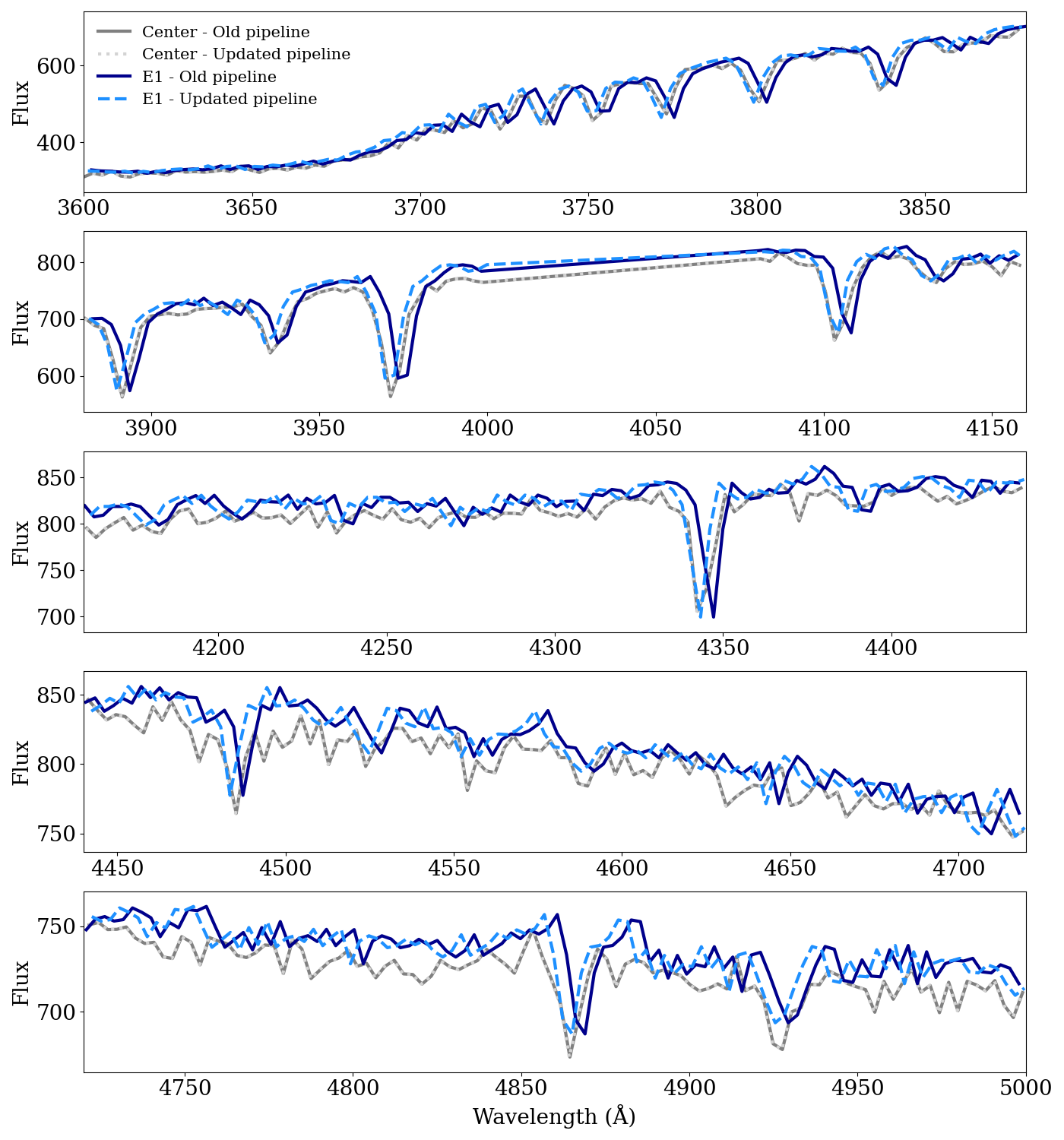}
      \caption{Comparison between the old and updated wavelength calibrations applied to G430L/4300 spectra of the source SK-67-207 (PID 16230), observed at both the nominal and E1 positions. Comparable results were obtained for the other sources with analogous nominal and E1 spectra available.}
         \label{fig:snaptest}
   \end{figure}

First, we made a test on one of the G430L/4300 wavecal datasets (dataset \texttt{oec63w010\_wav.fits}), calibrating it as a science exposure. 
In particular, the \texttt{\_wav.fits} file was first copied to a \texttt{\_raw.fits} file, and several header keywords\footnote{{\it WAVECAL} was set to the corresponding wavecal file, {\it FLUXCORR} was set to {\it OMIT}, {\it WAVECORR} to {\it PERFORM}, and {\it ASN\_MTYP} to {\it SCIENCE}.} were modified so that \texttt{calstis} would treat it as a science exposure. The {\it PROPAPER} keyword of the newly created raw wavecal file was then set to either \texttt{52X0.1} or \texttt{52X0.1E1}, so that the exposure would be interpreted as a science observation taken at the nominal or E1 position, respectively. Running \texttt{calstis} on these files produced the corresponding \texttt{\_flt.fits} files.
The SHIFTA1 value calculated with the original version of the pipeline is $\sim$3.3~pixel. The updated version returns the same value for the nominal position, but $\sim$4.7~pixel at the E1 position, corresponding to a difference of about 1.4~pixel between the two. These values are recorded in the resulting \texttt{\_flt.fits} files.
We then extracted one spectrum at the nominal position (512$\pm$150 pixel) and one at E1 (900$\pm$20 pixel) from the different \texttt{\_flt.fits} files using \texttt{stistools x1d}. Figure~\ref{fig:snaptest-wavecal} shows that the spectrum extracted at the nominal position (dashed red) is in very good agreement with the lamp template spectrum; the latter, degraded to the spectral resolution of G430L, is shown in magenta. 
By contrast, when the SHIFTA1 value from the original pipeline is used, the spectrum extracted at E1 is significantly shifted (dotted green line). When the updated SHIFTA1 value is used instead, the E1 spectrum (solid green) shows much better agreement.
Overall, the residual offset between the nominal-position and E1 spectra decreases from more than 1~pixel with the original pipeline to $\sim$0.1~pixel with the updated correction, below the target wavelength calibration accuracy threshold ($\sim$0.2~pixel).
This small residual offset of $\sim$0.1~pixel may reflect the rapid variation of the wavelength solution toward the detector edge, such that an extraction width of 40 pixels may already average over a non-negligible gradient, although this width helps preserve adequate signal-to-noise.
Finally, we applied both the original and updated pipeline versions to the other datasets from PID~16230 in which the same targets were observed at both the nominal and E1 positions. Figure~\ref{fig:snaptest} compares the spectra calibrated with the two pipeline versions for one representative target (note that the flux calibration step was not applied here), focusing on selected spectral regions containing prominent absorption features. The improved agreement between the nominal-position and E1 spectra confirms that the updated wavelength calibration provides a consistent correction when applied to science observations.

\begin{figure}[!h]
\centering
\includegraphics[width=.9\linewidth]{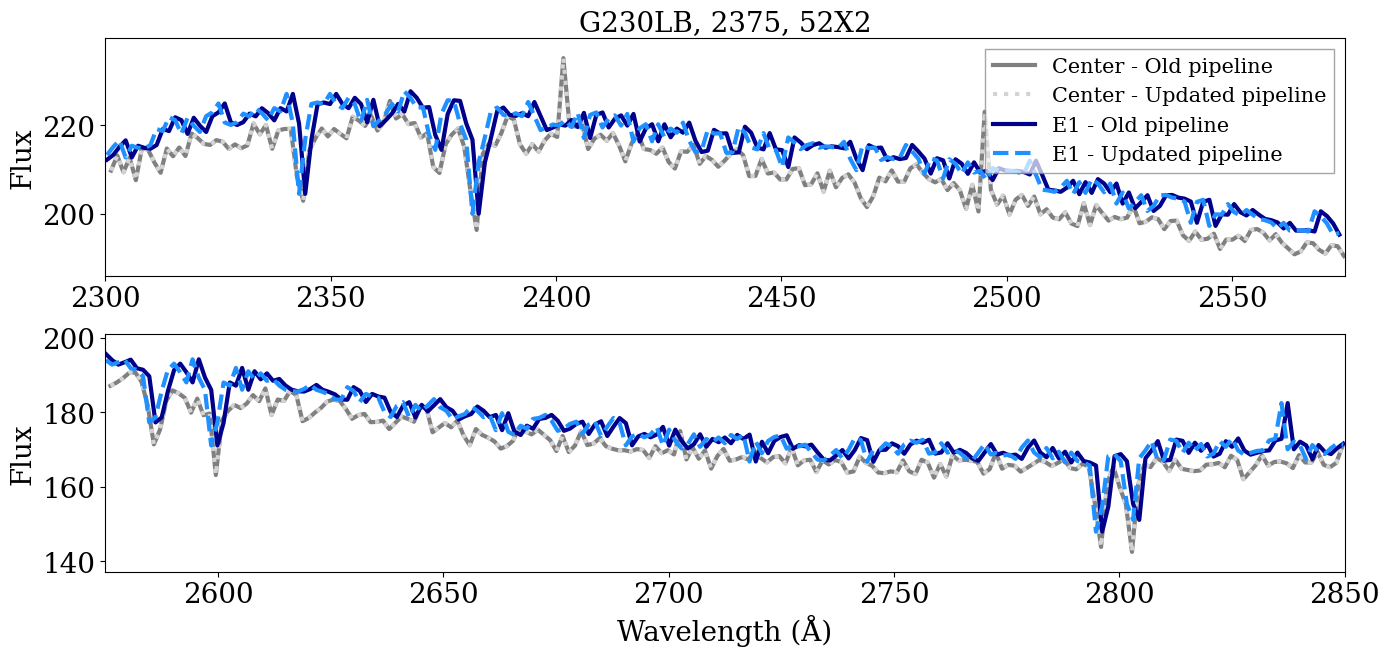}
\includegraphics[width=.9\linewidth]{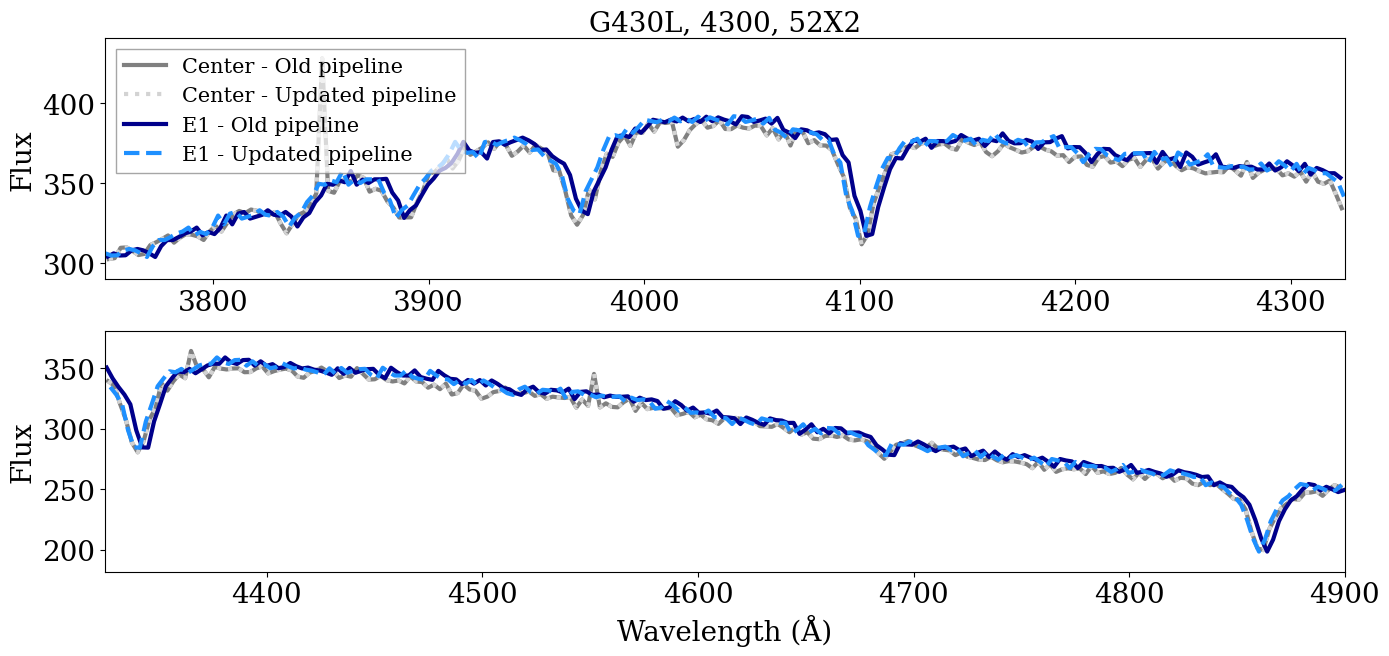}
\includegraphics[width=.9\linewidth]{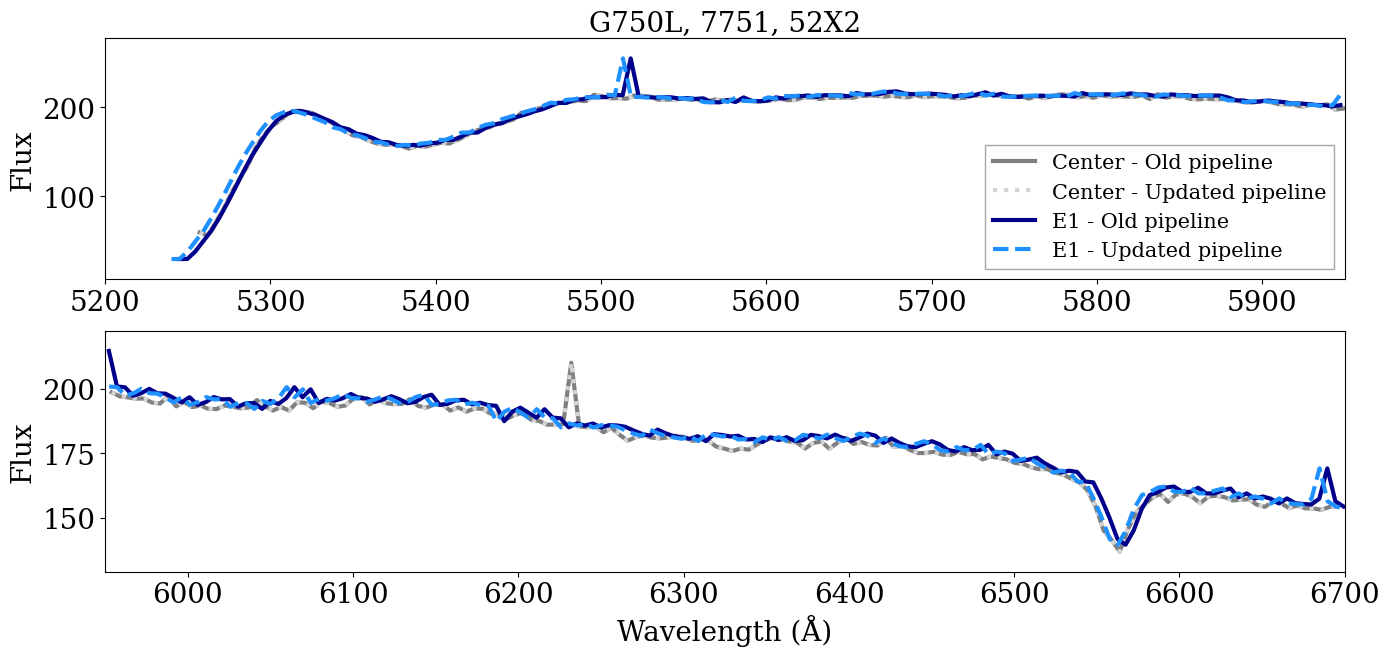}
\caption{Comparison between the original and updated wavelength calibration applied to representative G230LB/2375, G430L/4300, and G750L/7751 spectra of the standard white dwarf AGK+81D266 obtained at the nominal and E1 positions.}
\label{fig:tdstest1}
\end{figure}

A second validation test of the updated pipeline was performed using Cycle 27 STIS Spectroscopic Sensitivity and Focus Monitor data (PID 15747, visit L2). The target of this monitoring program is a standard white-dwarf characterized by smooth, well-modeled spectra and high signal-to-noise observations, used to track instrumental sensitivity and focus. Although these data are not ideal for assessing wavelength calibration accuracy, they provide a useful test case to verify that the updated pipeline does not introduce undesired side effects.
We ran both the original and updated pipelines on G230LB/2375, G430L/4300, and G750L/7751 datasets, including spectra acquired at both the nominal and E1 positions. 
Figures~\ref{fig:tdstest1} compare the spectra calibrated with the two pipeline versions for the three gratings (flux calibration step not applied here), focusing on selected spectral regions containing significant absorption lines. 
As expected, the results for the nominal position are unchanged, since the pipeline behavior in this case is identical to that of the previous version. For the E1 position, however, a clear wavelength shift is observed (see absorption line centroids), bringing the E1 and nominal spectra into closer agreement.

\begin{figure}[htbp]
\centering
  \includegraphics[width=.9\linewidth]{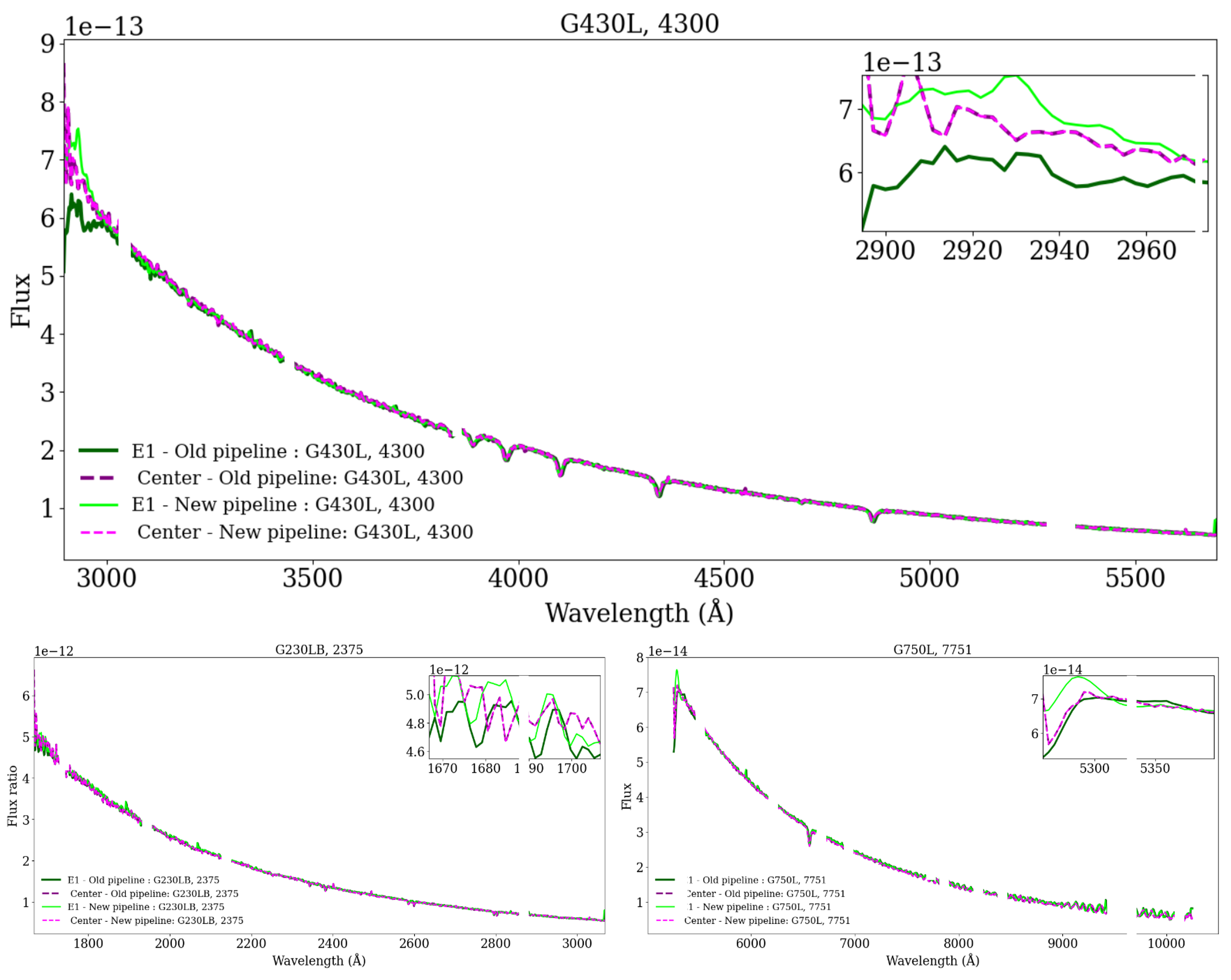}
  \caption{Comparison between the old and updated wavelength calibrations applied to flux-calibrated G430L (top), G230LB (bottom left) and G750L (bottom right) spectra obtained at the nominal and E1 positions of the standard white dwarf AGK+81D266, as indicated in the legend. The inset (on the upper right) highlights differences in both flux level and spectral shape at the shortest wavelengths, likely caused by the effect of the small wavelength shift on the sensitivity curves.}
     \label{fig:caveats1}
\end{figure}

\begin{figure}[htbp]
\centering
  \includegraphics[width=.9\linewidth]{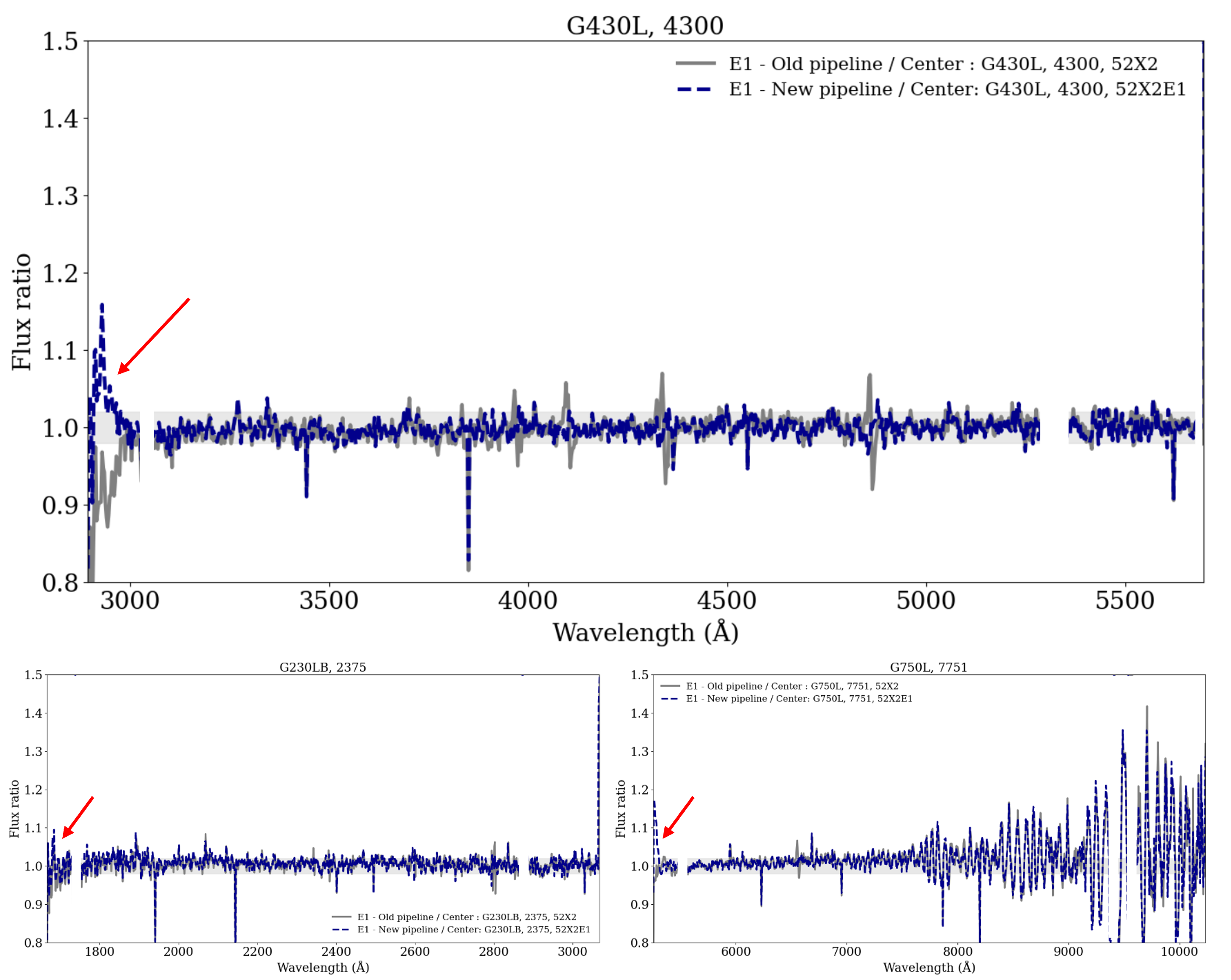}
  \caption{Ratio between the E1 and nominal spectra obtained with the standard pipeline (solid gray) and the updated pipeline (dashed blue) for the G430L (top), G230LB (bottom left), and G750L (bottom right; dominated by fringing beyond 7000~\AA) gratings. The flux-calibration differences  are most evident for G430L and G750L, below 3000~\AA\ and 5300~\AA, respectively, while no clear discrepancy is observed for G230LB (see red arrows).}
     \label{fig:caveats}
\end{figure}

During this second test, a difference in count rates was observed at the shortest wavelengths when comparing center and E1 spectra processed with the standard and updated pipelines.
This effect is likely related to the small wavelength shift affecting the sensitivity curves, raising the question of its possible impact on the flux calibration. Figure~\ref{fig:caveats1} shows the corresponding flux-calibrated spectra for the data presented in Figure~\ref{fig:tdstest1}, with a zoom-in of the shortest wavelengths to better highlight differences in both flux level and spectral shape.
To better quantify flux-calibration differences across the full wavelength range, Figure~\ref{fig:caveats} shows the ratio between the E1 and nominal spectra obtained with the standard (solid gray) and updated (dashed blue) pipelines for the G230LB, G430L, and G750L gratings. 
The effect is most evident for G430L and G750L (central and lower panels).
Overall, preliminary assessments suggest that this is not a major concern; however, the issue will continue to be monitored by the STIS team.

A final validation test was performed by cross-correlating wavecal 
exposures against the corresponding lamp reference files at both the nominal and E1 positions for a large sample of E1 datasets from the MAST archive\footnote{The validation sample included all available E1 datasets for G230MB, G430M, and G750M, and approximately 10\% of the E1 datasets for G230LB, G430L, and G750L. In total, we inspected $\sim 2,200$ E1 datasets. }.
\begin{figure}[!h]
\centering
  \includegraphics[width=1.\linewidth]{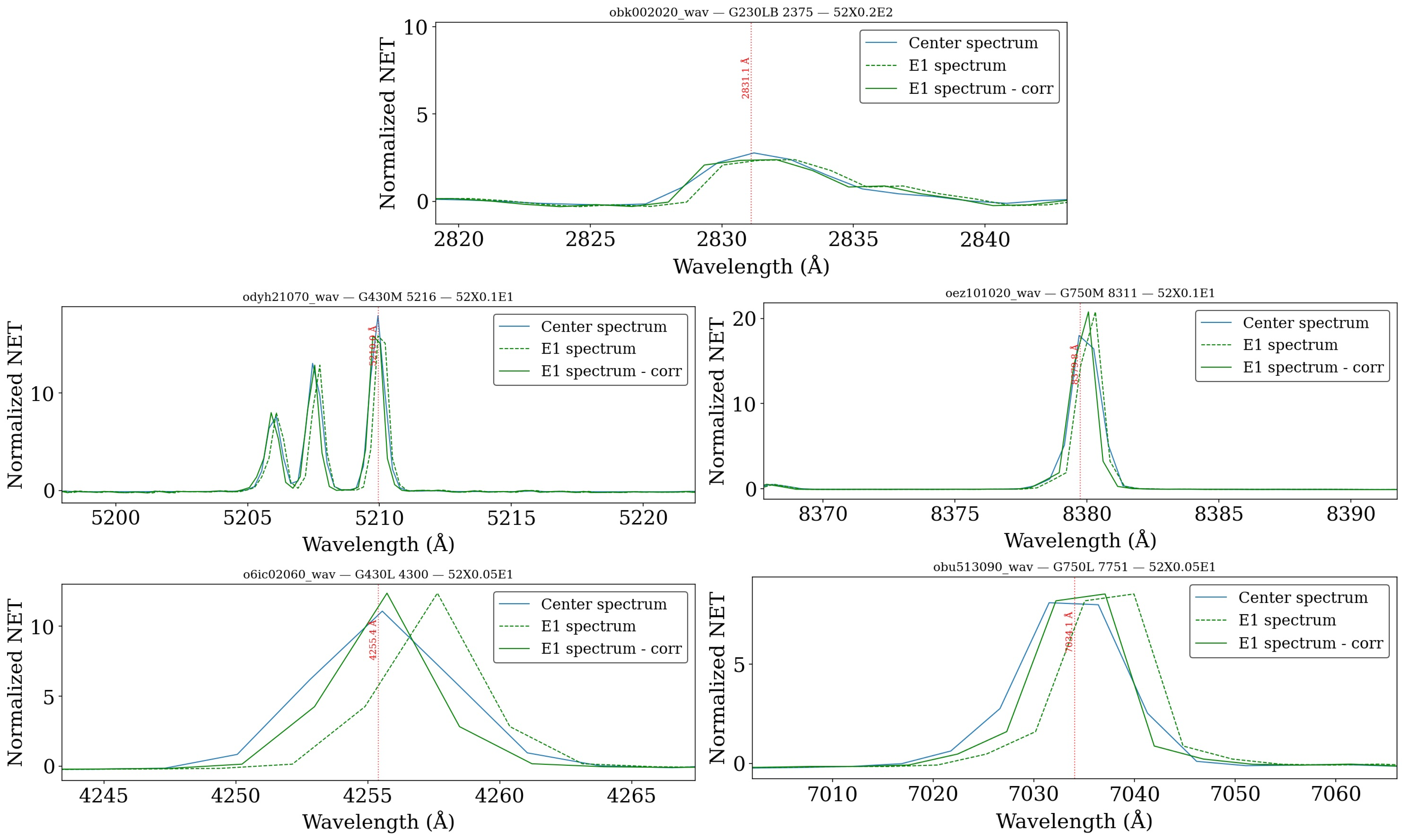}
  \caption{Comparison of extracted wavecal spectra at the detector center (solid blue) and at the E1 position before (dashed green) and after (solid green) applying the updated pipeline, for G230LB, G430M, G430L, G750M, and G750L. The improved agreement demonstrates the practical improvement in wavelength calibration accuracy for the vast majority ($>97$\%) of E1 datasets. This Figure is taken from the \href{https://www.stsci.edu/contents/news/stis-stans/may-2026-stan.html\#calstis_update}{May 2026 STIS STAN}.}
     \label{fig:stan}
\end{figure}
Considering the E1 usage (see Figure~\ref{fig:e1usage} in Appendix~\ref{app:e1-usage}), our results show that performing the cross-correlation at E1 improves the wavelength calibration accuracy for more than 97\% of the E1 datasets (Figure~\ref{fig:stan} shows a few examples). 
The main exceptions are G230MB observations, for which failures occur in 77\% of the datasets; a small number of failures were also found for G230LB/2375 ($\sim$7\%) and G430M ($\sim$6\%) datasets\footnote{Failures were found for most G230MB/1995, 2416, 2697, 2794, and 2836 datasets, and for a few G430M/3165,3680,3936,4194,4706,4961,5471; no failures found for G230MB/3115, G430M/3305, 3423, 3843, 4451, 5093, and 5216. No failures were found for G430L/4300, G750M/5734, 6094, 6252, 6581, 6768, 7283, 7795, 8311, 8561, 8825, and 9336, or G750L/7751.}. 
In these cases, the lower signal-to-noise ratio at the CCD edge makes the lamp lines more difficult to detect (especially if LINE instead of HITM1 is used), while cosmic-ray residuals can dominate the cross-correlation and lead to incorrect wavelength shifts.

\newpage
\subsection{Archive Reprocessing and User Guidance}\label{sec:guidance}
As presented in the \href{https://www.stsci.edu/contents/news/stis-stans/may-2026-stan.html\#calstis_update}{May 2026 STIS STAN}, the updated pipeline was used to reprocess the entire STIS archive in MAST.
The practical implications for users are summarized below, including the cases where additional checks of the wavelength calibration are recommended:
\begin{itemize}
    \item Users downloading data from MAST or reprocessing observations with \texttt{calstis} through the standard \texttt{stistools} workflows (with \texttt{calstis} version $\geq$~\texttt{3.5.0}) do not need to take any additional action, as the updated pipeline is already applied automatically. 
    \item Users who run the individual modules directly instead of using stistools must specify the -e flag when running cs4.e to ensure that the cross-correlation is performed at the E1/E2 position (e.g., {\it cs4.e -e rectified-file}). The \texttt{-e} flag is not required when running cs0.e.
    \item As said above, the majority of G230MB and a subset of G430M and G230LB datasets may show unreliable wavelength-calibration solutions because the wavecal cross-correlation at the E1/E2 position can fail, primarily due to low-S/N wavecal spectra and spurious features near the detector edges. When this occurs, the pipeline usually produces anomalous SHIFTA1 values (stored in the \texttt{\_flt.fits} or \texttt{\_crj.fits} file header), with shifts reaching tens of pixels instead of the expected few-pixel offsets. Users working with these modes are encouraged to inspect the wavelength calibration quality of their observations. In case of anomalous SHIFTA1 values, the jupyter notebook described in Appendix~\ref{app:wavecal-for-users} can be used to recalculate SHIFTA1.
    \item Observations obtained with non-default target placements (e.g., POS-TARG offsets or extended sources along the slit) may still show residual wavelength shifts due to CCD rotation and may require additional corrections. The jupyter notebook described in Appendix~\ref{app:wavecal-for-users} can be used in cases requiring high-precision wavelength calibration.
\end{itemize}

\section{Conclusions}\label{sec:conclusion}
In this work we investigated the wavelength calibration accuracy of STIS CCD data as a function of detector position and time, focusing in particular on spectra taken with E1 pseudo-apertures. 
Here we summarize our main findings:
\begin{itemize}
\item By comparing fitted lamp line centroids with laboratory wavelengths, we confirmed that the standard pipeline wavelength solution remains stable at the detector center but shows increasing offsets toward the CCD edges, consistent with previous studies, as a side-effect of the CCD rotation.
\item The scientific impact of the wavelength offsets can be significant for velocity-sensitive applications (e.g., spatially resolved kinematic and redshift/blueshift measurements). At typical optical wavelengths used for research studies (e.g., H$\alpha$), the measured shifts across the CCD for the L and M gratings could be erroneously converted into velocity offsets of $\sim$200~km~s$^{-1}$ and $\sim$20~km~s$^{-1}$, respectively, in case of extended sources or in velocity measurements of compact sources placed at the E1/E2 position. 
\item A test pipeline implementation demonstrated that performing the wavecal cross-correlation over a selected detector row range, instead of considering a summed 1-D spectrum collapsed over the full spatial direction, significantly improves wavelength calibration accuracy. Although this test implementation is not suitable for general release because it requires manual interaction, it motivated both an update to the \texttt{calstis} pipeline to improve wavelength calibration accuracy at the E1/E2 position and the development of a user notebook for a Row-Selected Cross-Correlation at any position on the CCD (see Appendix~\ref{app:wavecal-for-users}).
\item The updated pipeline applies a row-selected cross-correlation when E1/E2 pseudo-apertures are used, resulting in improved agreement between spectra extracted at nominal and edge positions while preserving the behavior at the nominal location. Validation tests on calibration and monitoring datasets confirm the expected wavelength correction and show no major adverse effects.
\item A small difference in count rates at the blue edge of some spectra was identified, likely related to the interaction between the wavelength shift and the sensitivity curves. The impact on flux calibration is currently under investigation; however, preliminary assessments indicate that this effect is not a major concern.
\item Observations obtained with some G230MB, a few G230LB and G430M settings or non-default target placements (e.g., POS-TARG offsets or extended sources along the slit) may still show residual wavelength shifts due to CCD rotation and may require additional corrections; see Section~\ref{sec:guidance} and Appendix~\ref{app:wavecal-for-users} for more details.
\end{itemize}
Overall, the updated pipeline provides a practical improvement in wavelength calibration accuracy for E1/E2 spectra, while maintaining backward compatibility.
The user notebook for row-selected cross-correlation (Appendix~\ref{app:wavecal-for-users}) provides a useful tool for cases in which wavelength calibration is required for spectra located neither at the nominal nor at the E1/E2 positions, or for spatially resolved CCD data where spectra need to be extracted at multiple positions along the slit, and a high level of wavelength accuracy is needed.

\vspace{-0.3cm}
\ssectionstar{Acknowledgements}
\vspace{-0.3cm}
We thank Maria Jesus Jimenez Donaire for her detailed review of this ISR, and Jacqueline Brown for her careful review of the Jupyter Notebook presented in Appendix~\ref{app:wavecal-for-users}. We also thank Daniel Stapleton for valuable discussions related to the archive reprocessing carried out as part of this work.

\vspace{-0.3cm}
\ssectionstar{Change History for STIS ISR 2026-04}\label{sec:History}
\vspace{-0.3cm}
Version 1: \ddmonthyyyy\today - Original Document 

\setlength{\bibsep}{0in}
\bibliography{mybib}

\newpage

\appendix

\vspace{-0.3cm}
\setcounter{section}{0}
\renewcommand{\thesection}{\Alph{section}}

\begin{figure}[!h]
    \centering
    \includegraphics[width=1.\linewidth]{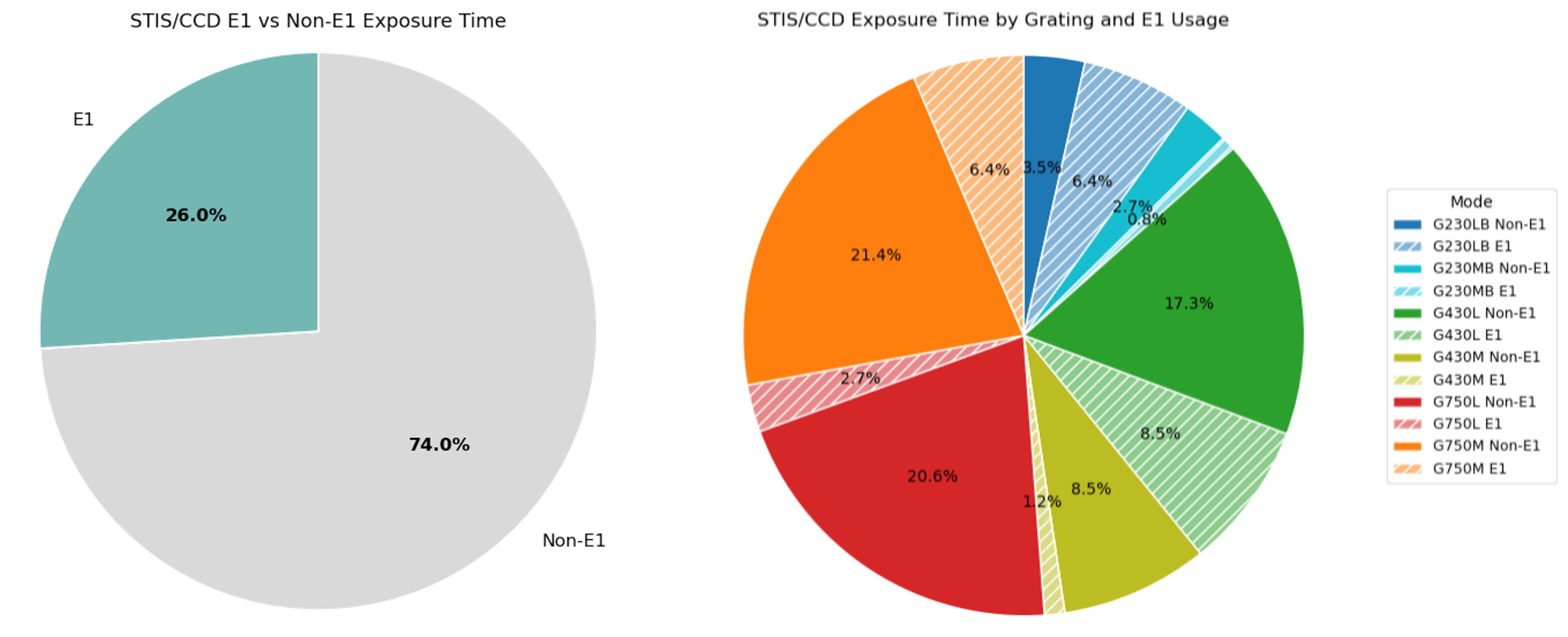}
      \caption{Usage of the E1 pseudo-aperture in STIS/CCD observations between 2000 and 2026. Left: approximately 26\% of the total STIS/CCD long-slit exposure time was obtained at E1. Right: breakdown by grating, showing that E1 observations account for a significant fraction of the exposure time in several commonly used modes. This motivates the improvements to the wavelength calibration at off-center detector positions presented in this ISR. }
    \label{fig:e1usage}
   \end{figure}

\appsection{Usage of E1 and E2 pseudo-apertures}\label{app:e1-usage}
Figure~\ref{fig:e1usage} shows the overall usage of the E1 pseudo-aperture in STIS CCD observations between 2000 and 2026, measured in exposure time (left panel), and its distribution across gratings (right panel).
The E2 pseudo-aperture has been used in about 20 programs using G750M and G750L so far.
   
\appsection{Wavecal-Based Evaluation of Wavelength Shifts}\label{app:wavecal-for-users}
The updated pipeline described in Section~\ref{sec:new-pipeline} improves the wavelength calibration accuracy for observations obtained at the E1/E2 positions, while the original pipeline already provides adequate accuracy for spectra extracted at the detector center. However, spectra extracted at arbitrary detector positions (use of POS-TARGs or extended sources) are not fully corrected by either the default or the updated wavelength calibration procedure. 
Also, as mentioned in Section~\ref{sec:guidance}, the cross-correlation at the E1 position can fail for some G230MB and a few G230LB and G430M datasets.
For this reason, we investigated how contemporaneous wavecal exposures can be used to improve the wavelength calibration accuracy in these cases, and provided a Jupyter Notebook\footnote{\href{https://spacetelescope.github.io/hst_notebooks/notebooks/STIS/e1_notebook/wavecal-cross-corr.html}{Improving Wavelength Calibration Accuracy Across the CCD}} implementing this procedure for users. 

For this notebook, we used wavecal data from PID 16230 (dataset \texttt{oec63w010\_wav.fits}, G430L/4300; same used in Section~\ref{sec:validation-tests}), although the specific dataset is not critical since the analysis would be equivalent for any wavecal observation.
To calibrate the wavecal exposure as a science exposure, we used the same method described in Section~\ref{sec:validation-tests}. 
We wavelength-calibrated the dataset using \texttt{stistools} by first processing the lamp wavecal files with \texttt{calstis} as if they were science data. 
The notebook includes two distinct examples:
\begin{itemize}
    \item The first illustrates a simple extraction of a spectrum at the nominal detector center and at an additional position near the top of the CCD, but not at E1 (row 750 in this example). This case is intended for point sources located away from the nominal center, but not exactly at the E1 position. Users can change this row value to match the position of their target in the science data. 
    \item The second applies the same procedure to 33 positions across the CCD, matching the \texttt{A2CENTER} values used in the STIS reference files. This case is useful for extended targets, or for users who want to inspect their wavecal spectra across the detector and determine how the measured shift changes with position. This may be also particularly helpful for G230MB, G430M, and G230LB datasets for which the cross-correlation at E1 can fail.
\end{itemize}
The extraction, cross-correlation, and correction steps described below are the same in both examples.

\begin{figure}[!h]
    \centering
    \includegraphics[width=1.\linewidth]{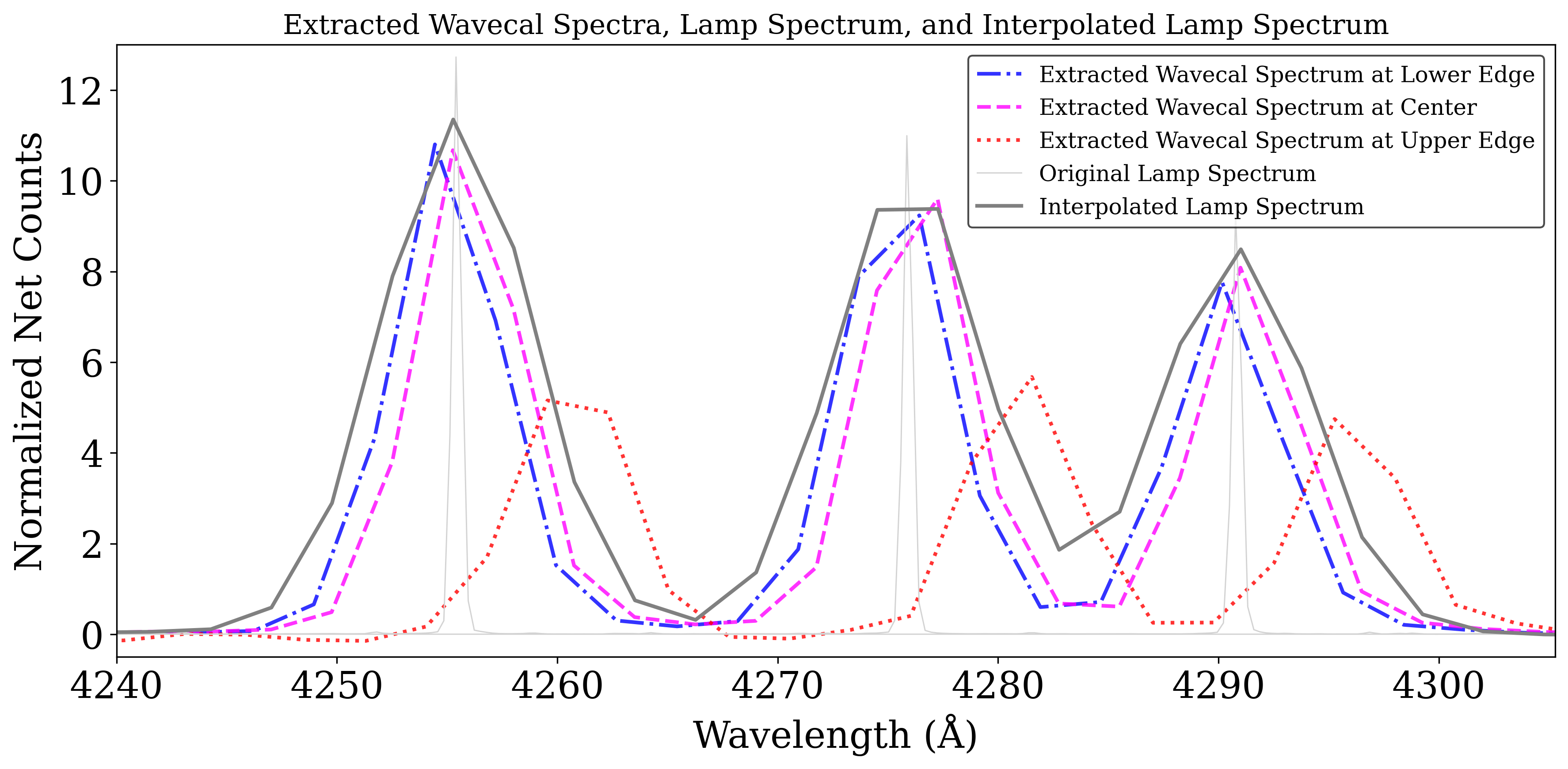}
    \caption{Comparison between wavecal spectra extracted near the detector bottom, center, and top (dash-dotted blue, dashed magenta, and dotted red), the reference lamp template (light gray), and the degraded and interpolated lamp template (solid gray). The latter is adjusted to match the local wavelength scale and resolution before performing the cross-correlation.}
\label{fig:wavecal-test-prep}
   \end{figure}

In particular, we extracted spectra from the calibrated lamp \texttt{\_flt} files, at different positions, using the \texttt{stistools.x1d} task (similar method as in Section~\ref{sec:comp-friedman} and Section~\ref{sec:test-pipeline}).
We retrieved the lamp template used by the wavelength calibration procedure \texttt{ l421050oo\_lmp.fits} from the STIS reference files\footnote{\url{https://hst-crds.stsci.edu/}}, and we degraded it and interpolated it to make it comparable with the extracted G430L spectra. 
Because the wavelength plate scale varies slightly across the CCD, the reference-lamp spectrum was degraded accordingly for each extraction.
Figure~\ref{fig:wavecal-test-prep} shows the comparison between three spectra extracted at the bottom, center and top of the detector (dash-dot blue, dashed magenta, dotted red), the reference lamp spectrum (light gray) and the degraded and interpolated reference lamp spectrum (gray). 
   
\begin{figure}[htbp]
    \centering
    \includegraphics[width=.93\linewidth]{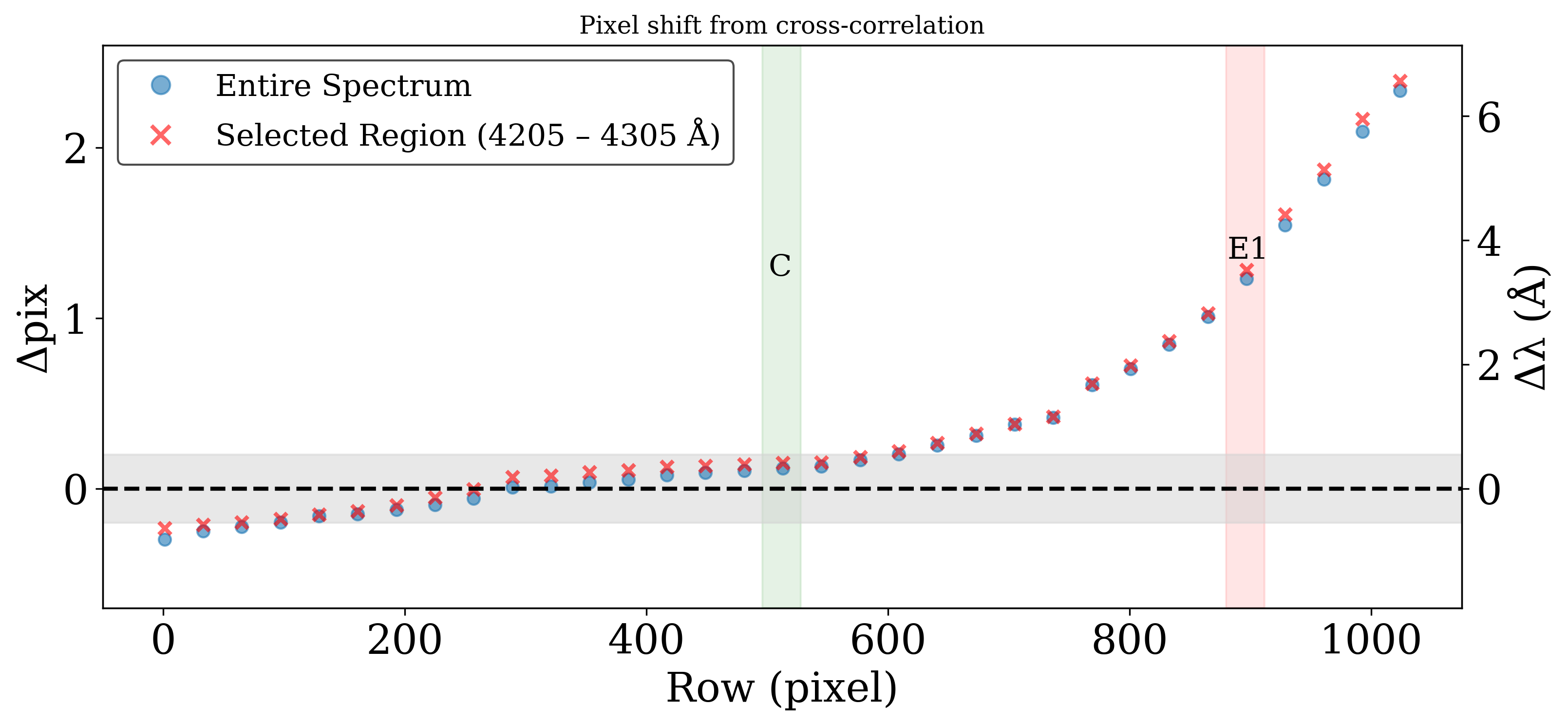}
    \caption{Results of the cross-correlation analysis applied to the full extracted wavecal spectra (circles) and to the selected region shown in the next figure (red crosses), looking very consistent; the lamp used in this example is the LINE lamp. The shifts are shown both in pixels and in \AA\ (left and right y-axes). The overall trend is consistent with that seen in Figure~\ref{fig:friedman-comparison} and Figure~\ref{fig:calprograms-test-pipeline}.}
     \label{fig:wavecal-test}
   \end{figure}

Then, instead of applying the Gaussian-fitting procedure described in Section~\ref{sec:comp-friedman} (which has the drawback of being time-consuming when multiple lines are used and whose results can be affected by line blending), we adopted a cross-correlation approach, similar to the method used by \texttt{calstis4} for wavelength calibration (see also \citetalias{mingozzi2026}).
In particular, we cross-correlated the degraded and interpolated lamp spectrum with each extracted spectra. 
The cross-correlation provides the shift between the spectra both in pixels and in \AA, where the latter is obtained by multiplying the pixel shift by the wavelength plate scale. 
For the second example, these shifts are shown in Figure~\ref{fig:wavecal-test}.
The resulting trend broadly follows that observed for G430L from dispersion solution monitor programs (bottom panel of Figure~\ref{fig:friedman-comparison} and Figure~\ref{fig:calprograms-test-pipeline}).

\begin{figure}[htbp]
    \centering
    \includegraphics[width=.93\linewidth]{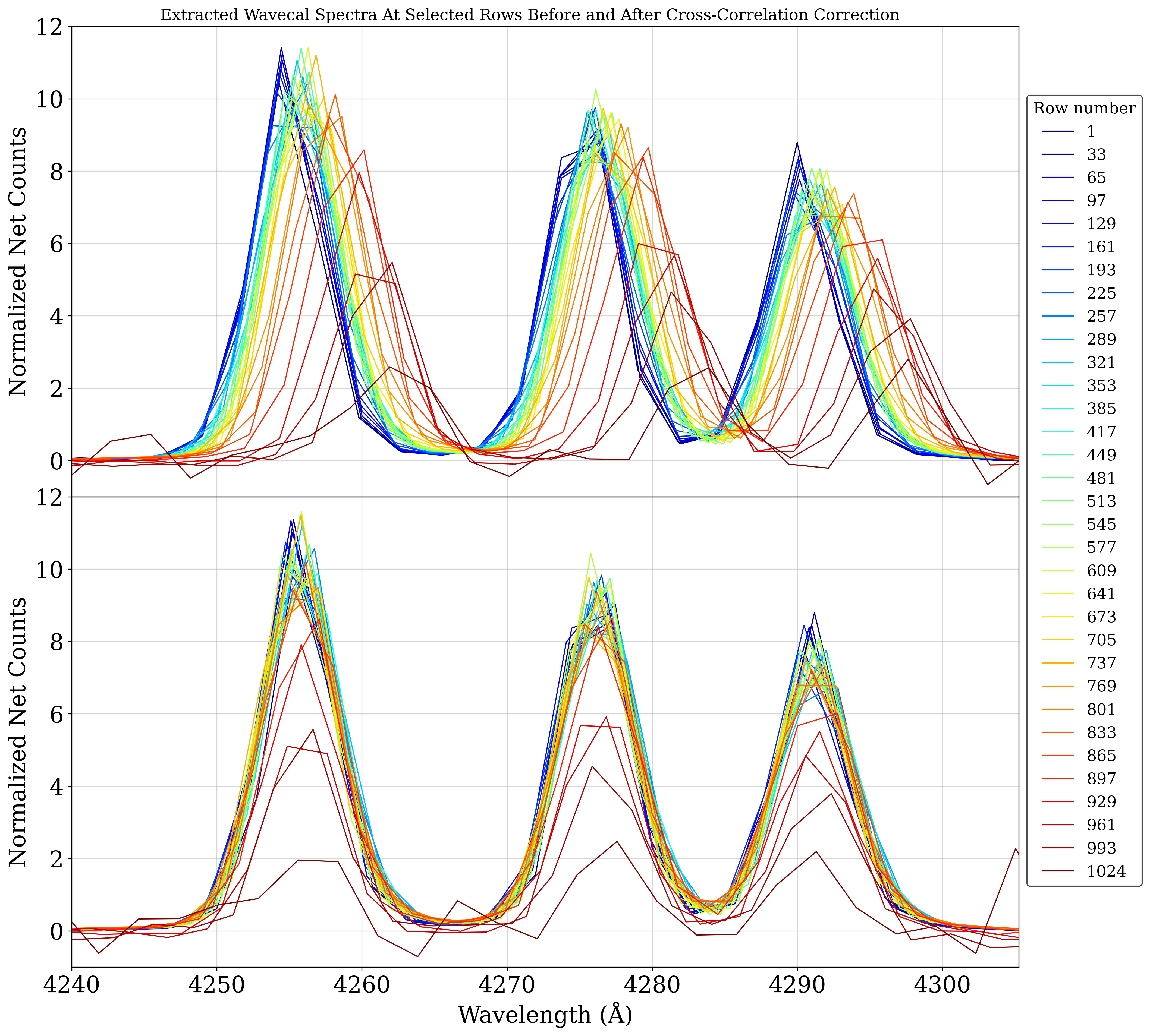}
    \caption{Upper panel: zoom on the extracted G430L wavecal spectra before applying the cross-correlation correction in wavelength. Colors indicate CCD row position (blue: bottom; red: top), as shown in the legend. Bottom panel: spectra after applying the residual wavelength shifts derived from the cross-correlation and shown in Figure~\ref{fig:wavecal-test}. This correction significantly improves the emission-line centroids' alignment across the detector.}
     \label{fig:wavecal-test-final}
   \end{figure}

The y-axis values shown in Figure~\ref{fig:wavecal-test} represent the residual corrections that must be applied to spectra extracted at the corresponding detector positions (x-axis) to improve the wavelength calibration accuracy. 
In particular, these correction can be applied by re-running \texttt{stistools.x1d} on the original \texttt{\_flt.fits} file with the \texttt{xoffset} keyword set to the corresponding pixel shift.
This is illustrated in Figure~\ref{fig:wavecal-test-final}, which shows a zoomed view of lamp spectra extracted at different positions along the CCD, as indicated by the color coding (blue corresponding to the bottom of the detector and red to the top). 
The top panel shows the original extraction, in which \texttt{stistools.x1d} applied the SHIFTA1 value stored in the header by the standard wavelength calibration procedure, while the bottom panel shows the result after correcting the extracted spectra.

Overall, this notebook has two main implications:
\begin{itemize}
    \item It demonstrates that the wavecal data can generally\footnote{As discussed in Section~\ref{sec:guidance} this procedure can fail for some G230MB, and a few G230LB and G430M datasets.} be used even near the edges of the CCD (including at E1), where the signal-to-noise ratio may be lower because lamp lines are fainter, but still sufficient for accurate measurements. This provides strong support for the \texttt{calstis} update presented in Sec~\ref{sec:new-pipeline}, which modifies \texttt{calstis4} to perform the cross-correlation around row~900 when observations are taken at E1.
    \item The updated version of \texttt{calstis} improves the wavelength calibration accuracy for spectra extracted at the E1 position. However, additional corrections may still be needed for sources that are not placed exactly at E1, or for sources that are extended across the CCD. For these cases, this notebook shows how to use the cross-correlation method described above to derive the appropriate wavelength corrections for the data.
\end{itemize}

\subsubsection*{Optional Step: Updating the DISPTAB and Reprocessing 2D Spectral Images}

The procedure described above derives position-dependent wavelength corrections that can be supplied to \texttt{stistools.x1d} through the \texttt{xoffset} parameter when extracting spectra at arbitrary detector positions. 
As an optional step, the notebook also shows how these offsets can instead be incorporated directly into a copy of the corresponding \texttt{DISPTAB} reference file. Since the measured offset varies with position on the CCD, but is approximately constant with wavelength at each extraction position \citepalias{mingozzi2026}, it can be treated as a position-dependent wavelength zero-point correction. Therefore, only the $A_0$ dispersion coefficient of the dispersion relation\footnote{See Section~3.4.8 of the STIS Data Handbook \citepalias{stisdhb} for further details.} needs to be updated at the corresponding \texttt{A2CENTER} positions, while the higher-order coefficients can be left unchanged.

\begin{figure}[!h]
    \centering
    \includegraphics[width=1.\linewidth]{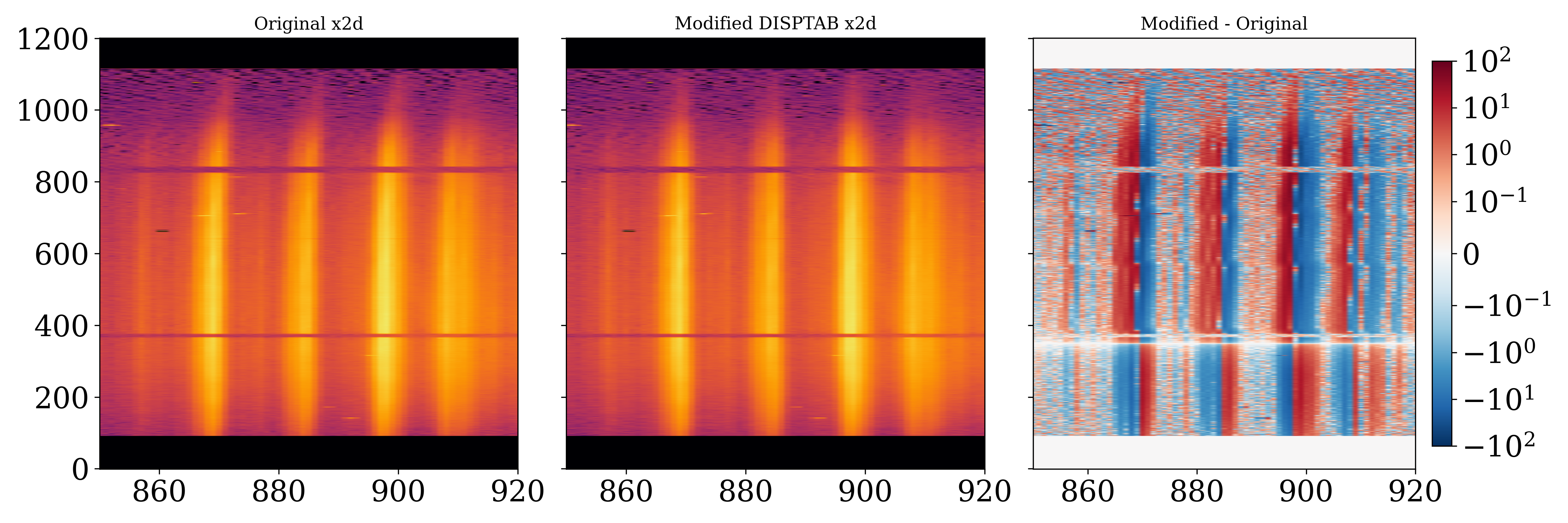}
    \includegraphics[width=.7\linewidth]{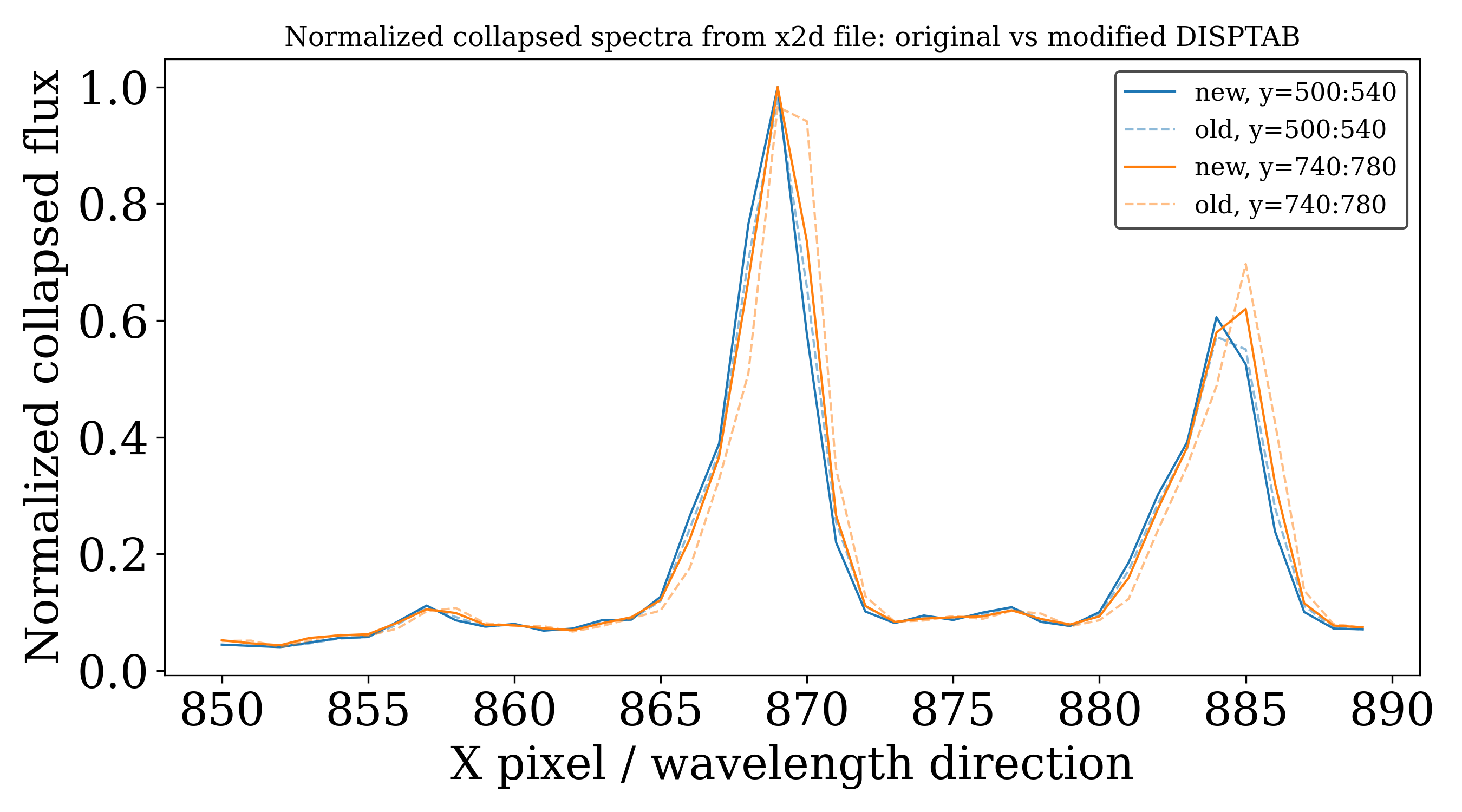}
    \caption{Upper panels: Comparison between the original and corrected \texttt{x2d} products after updating the \texttt{DISPTAB} reference file. The right panel shows the difference image (corrected minus original), highlighting the position-dependent shift applied along the dispersion direction. Bottom panel: Normalized profiles obtained by summing the flux in the original and corrected \texttt{x2d} products over selected row ranges. The displacement of the emission features confirms that the wavelength correction has been incorporated into the modified \texttt{DISPTAB}.}
     \label{fig:x2d}
   \end{figure}

As a result, \texttt{stistools.x2d} can generate corrected \texttt{x2d} products (useful in case of extended sources), and \texttt{stistools.x1d} can extract one-dimensional spectra without specifying an \texttt{xoffset} value.
To verify that the updated \texttt{DISPTAB} is correctly applied during the 2D rectification, we reprocessed the dataset \texttt{oec63w010} and compared the resulting \texttt{x2d} products. The upper panels in Figure~\ref{fig:x2d} show the original and corrected \texttt{x2d} images, while the bottom panel shows profiles obtained by collapsing the \texttt{x2d} flux over selected row ranges. The displacement of the emission features demonstrates that the wavelength correction has been correctly incorporated into the rectified 2D spectral image through the modified \texttt{DISPTAB}.
Because the derived offsets depend on the specific grating setting and observation, the modified \texttt{DISPTAB} is valid only for the dataset used to derive the correction and should not be applied to other observations.

\end{document}